**Title:** Instilling Doubts About Truth: Measuring the Impact of Tucker Carlson's Interview with Vladimir Putin using Machine Learning and Natural Language Processing

## Authors

Loni Hagen, PhD. lonihagen@usf.edu (Corresponding author)
School of Information, University of South Florida

Ly Dinh, PhD. lydinh@usf.edu
School of Information, University of South Florida

Golfo Alexopoulos, PhD. galexopo@usf.edu
School of Interdisciplinary Global Studies, University of South Florida

Lingyao Li, PhD. lingyaol@usf.edu
School of Information, University of South Florida

Diego Ford diegoford@usf.edu
School of Information, University of South Florida

Miyoung Chong, PhD. mc17@usf.edu
Department of Journalism and Digital Communication, University of South Florida

## Abstract

On February 7, 2024, Russian President Vladimir Putin gave a two-hour interview with conservative political commentator, Tucker Carlson. This study investigated the impact of the Carlson-Putin interview on the US X audience. We proposed a framework of social media impact using machine learning (ML) and natural language processing (NLP) by measuring changes in audience, structure, and content. Triangulation methods were used to validate the process and results. The interview had a considerable impact among segments of the American public: 1) the reach and engagement of far-right influencers increased after the interview, suggesting Kremlin narratives gained traction within these circles, 2) the communication structure became more vulnerable to disinformation spread after the interview, and 3) the public discourse changed from support for Ukraine funding to conversations about Putin, Russia, and the issue of "truth" or the veracity of Putin's claims. This research contributes to methods development for social media studies and aids scholars in analyzing how public opinion shapes policy debates. The Carlson-Putin interview sparked a broader discussion about truth-telling. Far from being muted, the broad impact of the interview appears considerable and poses challenges for foreign affairs leaders who depend on public support and buy-in when formulating national policy.

# 1 INTRODUCTION

On February 7, 2024, Russian President Vladimir Putin gave a two-hour interview with conservative political commentator and former FOX News host Tucker Carlson. The interview ran the following day "unedited" and "not behind a paywall" on Carlson's personal website and the X platform [8, 10]. This was Putin's first interview with the Western media since February 24, 2022, when he launched Russia's full-scale invasion of Ukraine.[1] Now in its third year, the Russia-Ukraine war represents the largest armed conflict in Europe since the end of WWII, with hundreds of thousands killed or injured and millions more displaced. Carlson's interview with the Russian president constituted a major media and international affairs event. As Paul Grenier noted at the time, "Two days after Tucker Carlson's talk with Russian President Vladimir Putin, the interview has racked up nearly 200 million views on Twitter alone. On a good night, MSNBC and CNN don't even get 2 million views" (Grenier, 2024). At the time of this writing, the interview has received over 20M views on YouTube, six times the viewership that Carlson's nightly show once received on FOX.

Given its popularity as a news distribution platform, the social media giant X, was likely chosen to release the interview and reach a broad US audience. According to a 2023 Pew survey, American X users use X for getting news (65%), and to keep up with politics and political issues (59%) [31]. The same survey revealed that although Twitter was predominantly used by Democrats initially, the acquisition by Elon Musk in 2022 resulted in the user ratio of X being more evenly divided between Democrats and Republicans.

Following the release of the interview, prominent Russian experts described Putin's intention "to return to some sort of dialogue" with U.S. conservatives [35], but they concluded that Putin was not successful in achieving this goal. Laruelle and Chrobak concluded that the interview did not gain significant traction among US audiences and was quickly overshadowed by the usual US news cycle. Even the US right-wing reaction to the interview was, the authors argued, soon "overshadowed by the regular U.S. news cycle" and was "relatively muted" [20]. However, this initial analysis of the interview's impact was based on the statements of select political leaders and news commentators.

In fact, Putin's interview with Tucker Carlson was broadly impactful if we apply a bottom-up approach and data-driven analysis. The interview produced a significant impact on the X platform, as many accounts engaged with various statements by Putin. For example, Laruelle and Chrobak mentioned Ben Shapiro's criticism of the interview, yet in responses to his X post, many disagreed with his critique [34]. The present work looks closely at the impact of an important social media event—Russian President Vladimir Putin's first major interview since he launched the largest land war in Europe in nearly 80 years.

We ask the following research question: *What was the impact of the Carlson-Putin interview on the US X audience?* This study investigated the impact of the interview by measuring changes in the audience, network structure, and content in communication on the X platform 48 hours before and after the interview. We also used machine learning (ML) and natural language processing (NLP) methodologies to measure communication on X, employing multiple triangulation methods.

[1] Putin's last interview with a major American media outlet came in October 2021 when he was interviewed by CNBC's Hadley Gamble. Guy Faulconbridge, "Kremlin confirms Putin gave interview to ex-Fox News host Tucker Carlson," Reuters (February 7, 2024).

The main contribution of this work includes methods development for social media studies. Our goal is to produce a framework for impact measurement and to propose triangulation methods when using a "data-driven" ML and NLP approach. Our findings obtained from triangulation of multiple methods show that the reach and engagement of far-right influencers increased after the interview, demonstrating the broad impact of the interview and the degree to which Kremlin narratives are shared among the far-right. In addition, the communication structure became more vulnerable to disinformation spread after the interview. Beyond social media studies, the present work is intended for foreign affairs scholars who understand how public opinion can influence policy debates. The Carlson-Putin interview sparked a broader discussion about truth-telling and revealed that many users debated and grappled with the veracity of Putin's claims. Far from being muted, the broad impact of the interview appears considerable, and poses challenges for foreign affairs leaders who depend on public support and buy-in when formulating national policy.

# 2 RELATED WORK

## 2.1 Social media and political communications

Social media has changed the way the public communicates, interacts, accesses information, and forms its political opinions. In traditional communication structures, a limited range of voices-- including political elites, institutional experts, and legacy media-- were responsible for disseminating information to the public in a one-to-many format [40]. Social media democratized the creation and distribution of information, producing a many-to-many structure of communication. This many-to-many communication structure enabled the immediate and explosive spread of information. Various entities can leverage this new ecosystem to attain diverse goals, from financial gain to political power. As a result, Tucker et al. (2017) assert that social media has become a battleground for political influence [40:48].

Social media is a technology that can democratize society by enabling individuals to challenge the status quo. It provides a platform for voices that are often excluded from mainstream media, including extremist groups. Social media, once considered a democratic tool, is now often viewed as a place where "more controversial ideas can go unchallenged." Algorithms may incentivize emotional messages and sensational headlines to promote the broad dissemination of these messages [40:52]. In addition, given the declining trust in institutions [1], people often turn to social media influencers for information. The vetting mechanism of information quality has weakened. Social media trolling can exert significant influence on civic discussion in ways that can shape public policy and legislation.

## 2.2 Measuring the impact of a social media event

Social media plays a crucial role in disseminating information that is crucial for opinion formation [5, 42, 44]. An act of seeing news items on X could be significant in shaping an individual's opinion, as it influences an audience's process of acquiring and weighting information [32]. [5] concluded that social media influences the behavior of its users, specifically in terms of voting behaviors on Facebook. However, it remains uncertain whether the changes in voting behavior are due to peer pressure rather than shifts in political beliefs [48].

Despite many efforts, measuring the impact of a social media event is still challenging. Harold D. Lasswell's act of communication theory is a well-known framework to describe the process of communication by examining who says what, in which channel, to whom, and with what effect [21]. This framework helps methodologists to measure media effect[2] by analyzing the communicator ("who"), the content ("says what"), and the audience ("to whom"). However, measuring the impact of an event on social media presents additional challenges, such as overlapping discussions from multiple events happening at the same time (Belcastro et al., 2021), and cascading effects where secondary events triggered from the original event arose (Deng et al., 2020). As a result, it is difficult to isolate the effect of a single event. In addition, the fast-paced nature of social media means that trends and influences can change quickly, making it difficult to capture and analyze events in real-time. A universal set of metrics for measuring the impact of an event on social media communications is also lacking [39]. This highlights an important limitation in using the act of communication theory, a post-World War II communication theory, for social media analysis: it cannot explain how people's interactions aggregate into large-scale patterns.

Social network analysis enables investigating individuals and their social environment to explain social phenomena [37]. A core principle of the field is that structure matters [7:893]. Social network analysis provides statistical measures to capture individual (node-) and structural (network- level) characteristics of communication networks.

**Node-level statistical measures.** In an X communication network, an account can attract many followers who like, share, or reply to its posts. Such a popular social media account is often called an influencer. Influencers are often located in the center of information flow because they can "tap into larger stores of useful political information" [18:276] or can function as a gate keeper. Such "prestige" of a node can be measured using a centrality or PageRank algorithm. Centrality can measure the potential power of an actor in increasing or slowing down information flow in a community [7].

**Network-level statistical measures**. As a social media platform, X generates its impact through interactions between users. User interactions create measurable patterns in network-level structure, called network properties. Network properties matter because, even with members with similar skill sets, how these members are connected to each other in the network influences outcomes. For example, certain types of community structures are known to be more advantageous for the contagion of ideas, which in turn could influence collective action [14]. This idea is based on Burt's structural holes theory of social capital. Structural holes refer to gaps between non-redundant contacts within a social network. These gaps exist when a node is connected with multiple groups that do not directly interact with each other. Pink nodes in Figure 1 are bridge nodes. Node A has more structural holes than Node B. The higher level of structural holes of node A indicates many nodes are not yet connected, but the information A may receive from the bridge nodes is more diverse than B. Therefore, A can have early access to new information and thereby function as "an amplifier for creativity." A lack of structural holes in node B (nodes with stronger ties) indicates that a node and nodes' contacts share similar ideas. This type of community structure is beneficial for the unified coordination of certain groups, such as unions and political parties [7].

---

[2] Media scholars investigated media effect, which measures rather short-term and direct changes to human behaviors. In this study, we use "impact" consistently to stress broader or indirect changes triggered by the interview.

To capture these theories with quantifiable measures, network scholars have developed statistical measures, observing patterns in path and geodesic (shortest path) distances[37].

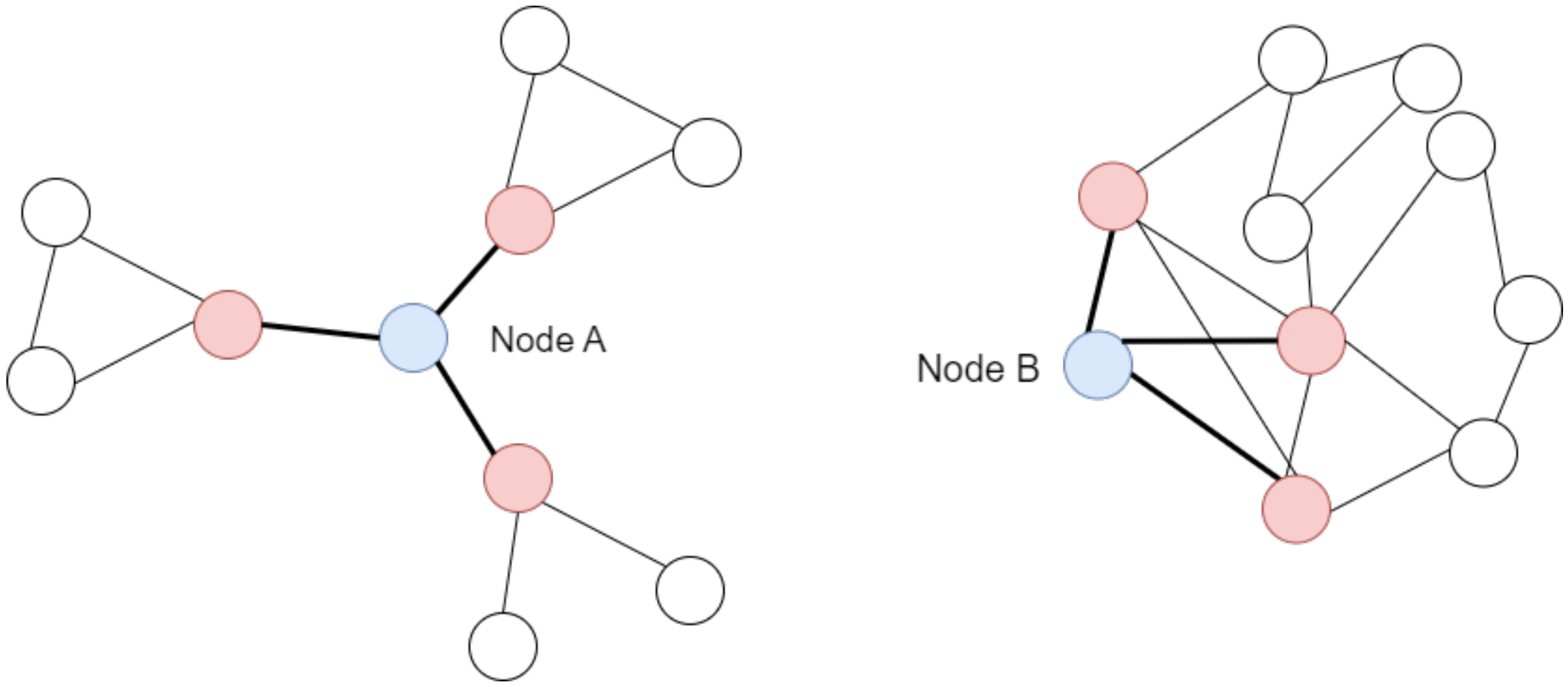

Figure 1. Node A has more structural holes than B (Note: This image is a reconstruction of [6])

**Natural language processing to measure text content.** Social media produces large volumes of written texts. Written text offers social scientists an effective window for interpreting actors' perception of themselves and the world around them [24, 45]. Using ML and NLP, social media texts have been mined to discover hidden patterns from big data. ML and NLP, supported by artificial intelligence (AI) more recently, have made promising contributions to the development of knowledge "in poorly understood, or complex" areas of diverse social phenomenon, and in "describing" new and unknown phenomenon [13]. NLP is used to automatically extract topics or identify sentiment addressed in social media, or to classify user characteristics or content categories [2, 9, 28, 30, 47].

While AI-driven NLP enables large-scale social media analysis, it raises concerns among scholars due to its reliance on black-box algorithms, which complicates validation. For example, Roberts et al. (2021) found methodological flaws, including poor application of ML methodology, in studies using machine learning [33]. The reliance on black-box algorithms has made AI interpretation and validation difficult in both technical and social science research. Triangulation could provide a method to produce valid results from computational methods.

## 2.3 Triangulation

In social science research, triangulation refers to the use of more than one research approach [16, 41:243]. The objective is to reduce biases, and to increase confidence in findings "through the confirmation of a proposition using two or more independent measures" [16:98]. Triangulation was originally developed in the qualitative research tradition in the 1950s to avoid biases by depending on a single method. In a sense that two or more methods are used, triangulation could be considered similar to mixed methods (typically by combining qualitative and quantitative research methods). However, triangulation could comprise wider scope since it includes data, investigator, method, and theory triangulation.

*Data triangulation* involves using data from different times and platforms. This type of approach has been used in previous studies to enhance the robustness and generalizability of the work, without using the term triangulation [2]. *Investigator triangulation* involves multiple researchers in the process of collecting and analyzing data. *Methodological triangulation*, the most common type of triangulation, uses diverse methods and multiple measures to approach the same research question, thus improving the work's validity. The use of more than one measure can help researchers discern "method variance" associated with any single measure [26:204]. The use of multiple measures of different types, all with varying weaknesses, can help countervail the weaknesses of any specific research method. *Theory triangulation* involves using diverse theories to approach the same research question.

The three primary anticipated outcomes of employing triangulation are as follows: 1) the findings may converge, leading to a unified conclusion; 2) the findings might complement each other, thereby enhancing the individual results; and 3) the findings could be "divergent or contradictory" [16].

# 3 METHODS

In this chapter, we briefly introduce the case and discuss the research methods in detail.

## 3.1 The case introduction

The present study examines a major newsworthy event: Tucker Carlson's interview with Russian President Vladimir Putin. The interview was aired on Calson's website and on the X platform on February 8, 2024. We chose this particular case because the Putin-Carlson interview represented a significant story in the news cycle. It constituted the first wartime interview with the leader of a global nuclear power, whose invasion of Ukraine launched the first major ground war in Europe since the end of WWII. We recognized that this important and unusual event sparked substantial discussion on social media. We aimed to delve further to measure and evaluate the real impact of the interview on X platform users with ML and NLP using a triangulation approach.

## 3.2 Triangulation

The following subsections discuss our triangulation approach (investigator-, data-, and methodological- triangulation) in detail (Figure 2 shows our overall strategy). Each step of triangulation constitute not linear, but rather cyclical (iterative) processes [13].

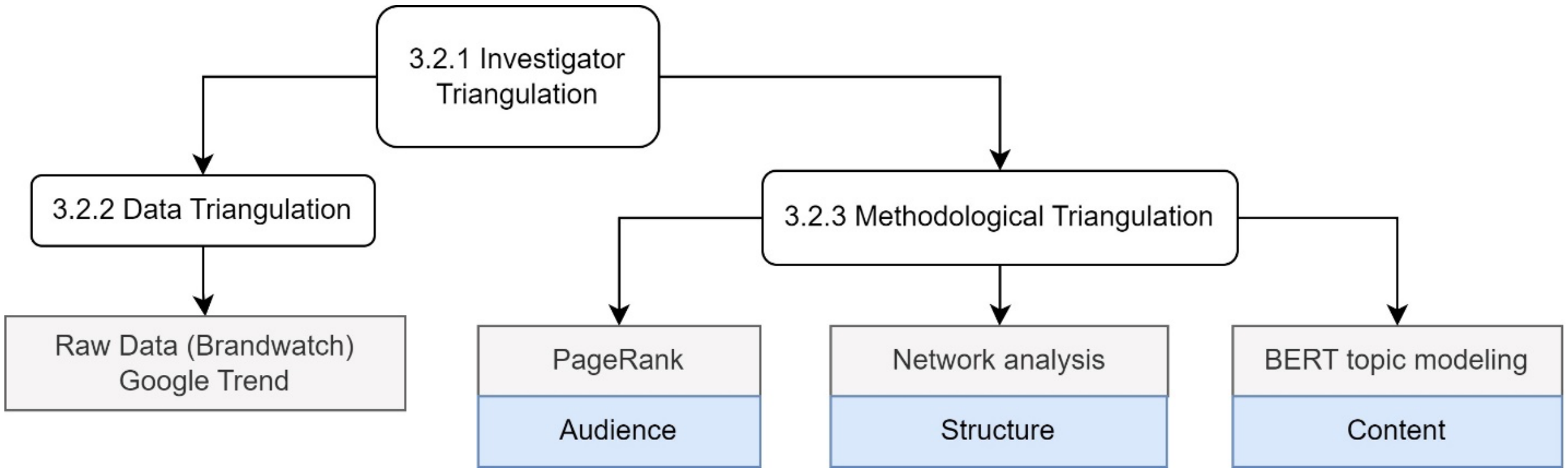

Figure 2. Triangulation strategy.

### 3.2.1 Investigator Triangulation

The study's interdisciplinary research process involved a domain expert in the field of Russian studies and a technology expert in the field of data science. These scholars took part in every step of the investigation, from data collection through interpretation and narrative construction, to minimize biases that could arise from each individual's disciplinary lens. During the interpretation phase, the two experts independently conducted analysis and met to generate a consensus. This method helped to reduce investigator biases and validate the study's interpretation.

### 3.2.2 Data Triangulation and Collection

X data was collected using Brandwatch, a social media analytics and listening tool, that provides historical social media data collection options for researchers. First, we developed a list of keywords to capture the general discourse about Russia and Ukraine as well as specific narratives Putin communicated during the two-hour interview. Both domain and technical experts worked together to determine the keywords. The final keywords were *("ukraine" OR "russia" OR "lenin" OR "poland" OR "nato" OR "putin" OR "Hitler" OR "WWII").*

Second, the time-range of data collection was determined through data triangulation, that is, iterative investigations using two different datasets (Google trends and the raw X data collected from Brandwatch). Determining the time range was both highly important and challenging: A long date range might include numerous irrelevant events, therefore noises, while a short date range might not capture the full impact of the event. An initial investigation using the two datasets with different time ranges revealed that 48 hours before and after the interview represented the optimal cut-off point. Within this range, the frequency of posts came down to the same level as before the event, while also capturing the impact of the interview before the interruption of other social events.

Third, geographical boundaries were determined to be in US and English posts only, to observe the impact of the interview on the US audience. We validated Brandwatch's location data by comparing it with manual coding, achieving 94% agreement (**Supplement 1**). We then used this location data. Of the total posts (N=1,339,274) during the time range of Feb 6$^{th}$ at 6 PM to Feb 10 11:59 PM EST of 2024, 46% (N=617,793) included a location tag, with 58% of these from the US. Consequently, 27% of all posts (N=361,018) that originated from the US were used for the initial analysis (Descriptive statistics of the data are in Supplement Table 1).

In preparation for network analysis, node and edge lists were created. By using repost and reply as the edge, we expect to capture the level of "repetition" and "amplification" of the content. For natural language processing, we determined to use the original posts only to capture naturally rising topics by the authors of the original posts. By using different relationship types, repetition vs. original contents, as input data for different analyses, we expect to triangulate the data and findings.

### 3.2.3 Methodological Triangulation

We consider the release of the interview as an intervention to X communication. To determine the impact of this intervention, we adhered to the social media impact framework described in the prior section, measuring the audience, structure, and content of social media communication (Figure 2). The interview's impact was assessed by comparing the changes in measures before and after the interview. The entire script for this work is available in this link: https://github.com/Big-Data-Analytics-Lab-USF/Putin_Interview/tree/main

#### *3.2.3.1 Network analysis for audience and structure analysis*

When node A shared a post created by B, there is a directed edge A ß B, following the information flow and therefore the directionality of influence. The "before" data included 40,044 nodes and 57,800 edges. The "after" data included 56,053 nodes and 87,407 edges. We conducted node-level statistical analysis to investigate "audience" and network-level statistical analysis to investigate "network structure." Gephi 0.10, an open source software for network analysis [3], was used to measure network statistics using a personal computer with Intel Core i7 and 32 GB RAM.

**Node-level statistical analysis to investigate audience behavior**: In our case study, the communicator was Putin and X users were the audience. X users, as audience, usually can take on particular roles such as leaders or followers depending on their behavior and action [22:9]. Social media leaders are often called influencers or opinion leaders, since they can garner substantial influence on their audience, accumulated through their engaging content creation or their ability to authentically connect with their audience. This study measured influencers through 1) simple descriptive analyses (i.e., number of followers and hashtags) and 2) an algorithmic approach. For the algorithmic approach to detect influencers, the PageRank algorithm was used. PageRank is an algorithm originally developed by Larry Page and Sergey Brin of Google (Page et al, 1999). The core idea behind PageRank is to determine the importance of a web page based on the number and quality of links directing to it. PageRank is popularly used to find influencers in social media (Hagen et al, 2018).

**Network-level statistical measures to capture network structures:** Network structure captures behaviors among related audiences and the spread of information in communities. The average degree, network diameter, and modularity were used to capture the network structure. **Average degree** is the mean number of connections per node in a community. When comparing the structure of two or more communities, the community with a higher average degree indicates more connections among nodes. **Network diameter** refers to the longest of the shortest paths between any two pairs of nodes. A smaller diameter indicates quicker information spread. **Modularity** is a tool for community detection, comparing the density of connections within a community against those between a community. A community is formed when a group of nodes is frequently connected to one another, as compared to the density between communities. [4]'s modularity optimization algorithm was used for community detection and [19] is used for modularity resolution. The default value was used, except in the case of the modularity resolution (set as 2) to create a manageable number of communities.

#### *3.2.3.2 Natural language processing*

Our dataset comprises 14% of all X posts that include one or more keywords present in the query. We used available metadata to reconstruct original posts that were not included but had existing reposts in the data. This approach allowed us to enhance the number of original posts for NLP tasks. As a result, the "before" data included 75,873 posts, and the "after" data included 114,837 posts (Supplement, Table 2).

**BERT Topic Modeling.** Topic modeling offers a method for identifying semantic themes from textual data [43]. It has been widely used to extract high-level insights from a large volume of textual data, such as social media posts, news articles, and other text sources [38, 46]. To identify topics in our X dataset, we utilized the BERTopic pipeline [15], which applies BERT embeddings to capture nuanced, context-aware representations of text content. While conventional statistical topic modeling methods like Latent Dirichlet Allocation (LDA) have been used extensively [17], we chose BERTopic for its ability to incorporate word meanings within context—a critical feature for accurately understanding the public discourse concerning the Carlson-Putin interview. Prior research has also demonstrated that BERTopic not only surpasses LDA and Top2Vec but also achieves comparable performance to prompt-based large language models (LLMs) in identifying topics from online data [12]. In this study, we added investigator triangulation to the topic modeling process (see Figure 3). For example, interpretability of the multiple BERT embedding models ("all-MiniLM-L6-v2" and "distilbert-base-cased") were tested, which revealed that "all-MiniLM-L6-v2" produced more coherent and interpretable results.

Once each post was transformed into a vector representation, we performed dimensionality reduction to manage the high dimensionality of BERT-generated vectors, which could otherwise be computationally intensive and complex to process. We incorporated Uniform Manifold Approximation and Projection (UMAP) within the BERTopic pipeline to achieve this [27]. UMAP can help reduce dimensions while retaining both the local and global structures of the data, optimizing it for clustering and making the results more interpretable.

Since the initial topic modeling results produce over 200 topics, we applied the elbow method to select a smaller number of topics feasible for human interpretation [25, 36]. The elbow method, based on the Within-Cluster Sum of Squares (WCSS) across a range of cluster counts, indicated around 50 clusters (Supplement, Figure 1).

The final stage of our topic modeling pipeline involved representing each clustered topic. To achieve this, we applied a count vectorization technique known as class-based Term Frequency-Inverse Document Frequency (c-TF-IDF) [15]. This method can help efficiently tokenize topics, allowing us to identify key terms and representative documents within each cluster. Using these representative terms and documents, we manually categorized each of the 50 topics into umbrella topics (UTs).

The 50 topics were manually investigated by two annotators (one domain specialist and one technical specialist). Two annotators were provided with (1) 50 topic words, (2) three representative documents per each topic, and (3) temporal graph of each topic, then were asked to independently interpret the 50 topics to extract UTs (Appendix I includes 50 topics and representative posts). The two annotators met to discuss and agreed to extract four UTs. The annotators closely examined the 50 topic words as well as the representative documents to determine whether the topics could be grouped under broader themes. The annotators agreed with four UTs, namely, 1) Truth telling, 2) Putin and Russia, 3) Ukraine War, 4) US and West. Relevant topics are assigned to these four UTs (Appendix II presents UTs and topic IDs assigned to each of the UTs).

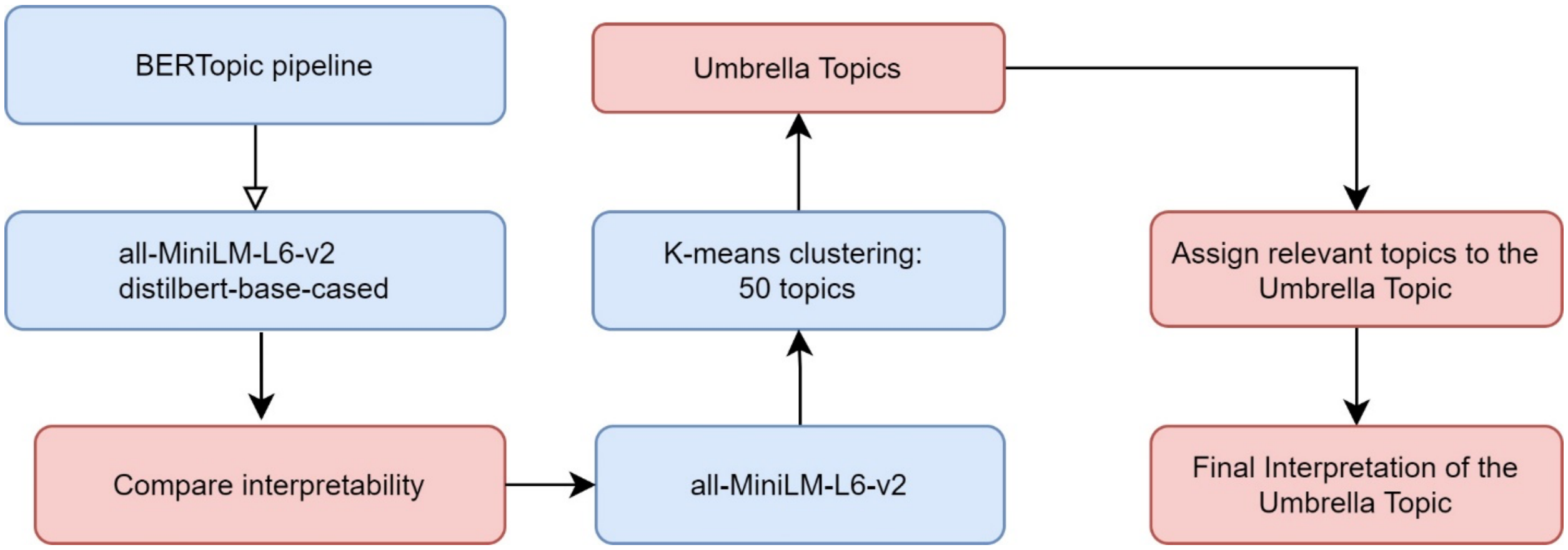

Figure 3. BERT topic modeling workflow (Note: the red color indicates investigator triangulation, and the blue color indicates computational work).

# 4 RESULTS

We present measured changes in audience, structure, and content observed by comparing X users before and after the interview.

## 4.1 Audience

We used descriptive analysis to understand the overall audience and PageRank to analyze influencers. Descriptive analysis reveals that the interview sharply increased user engagement with the topics featured in the conversation and dramatically enhanced the visibility of Tucker Carlson and Vladimir Putin. The influencer analysis shows that the top influencers with a far-right political orientation gained visibility after the interview.

### 4.1.1 Number of authors

After the interview, the peak frequency of X posts was close to 200% higher than the prior peak, showing a notable increase in audience engagement (Supplement Figure 2). The unique number of authors who created new posts increased about 80% and the number of original posts increased about 76% after the interview (Supplement Figure 3). Before the interview, popular hashtags included #supportukraine and #demvoice1, signaling the prominence of accounts that supported Ukraine's war effort and opposed Putin. Popular hashtags surged over ten folds after the interview. Notably, #tuckercarlson rose from 152 to 8536 mentions, and #putin increased from 64 to 7372, indicating significant public exposure for Tucker Carlson and Vladimir Putin (Table 1). The popular hashtags shows that the interview also intensified engagement with topics that the two men discussed — Russia, Ukraine, and NATO.

### 4.1.2 Influencers

The top 10 influencers were extracted using PageRank. Our analysis of the top 10 influencers revealed that: 1) right-wing actors heavily engaged both before and after the interview, and 2) new right-wing actors entered the top influencer group post-interview, indicating notable impacts on audience engagement (Table 2).

Table 1. Descriptive statistics

| | Authors (Original post) | Authors (Original + Repost) | Original Posts | Repost | URLs | Popular hashtags | Freq |
|---|---|---|---|---|---|---|---|
| Be-fore | 2,363 | 46,000 | 2,774 | 57,629 | 8,682 | #ukraine<br>#sup-<br>portukraine<br>#tuckercarlson<br>#tuckerputin<br>#putin<br>#demvoice1 | 626<br>231<br>152<br>80<br>64<br>63 |
| After | 4,246 (▲80%) | 64,122 (▲39%) | 4,893 (▲76%) | 88,192 (▲53%) | 13,498 (▲56%) | #tuckercarson<br>#putin<br>#ukraine<br>#russia<br>#tukerputin<br>#nato | 8,536<br>7,372<br>5,299<br>2,718<br>1,909<br>1,736 |

Table 2. Top 10 influencers and their political leanings before and after the interview

| Rank | Before | Political Ideology | After | Political Ideology |
|---|---|---|---|---|
| 1 | TuckerCarlson | Right-wing | TuckerCarlson | Right-wing |
| 2 | WarClandestine | Right-wing | WarClandestine | Right-wing |
| 3 | simonateba | Right-wing | **EndWokeness** | Right-wing |
| 4 | RealAlexJones | Right-wing | **RonFilipkowski** | Left-wing |
| 5 | bennyjohnson | Right-wing | bennyjohnson | Right-wing |
| 6 | MattWallace888 | Right-wing | **CollinRugg** | Right-wing |
| 7 | VivekGRamaswamy | Right-wing | **VigilantFox** | Right-wing |
| 8 | charliekirk11 | Right-wing | **catturd2** | Right-wing |
| 9 | seanmdav | Right-wing | **BasedMikeLee** | Right-wing |
| 10 | DavidSacks | Right-wing | charliekirk11 | Right-wing |

Note: accounts in bold are those that increased to a higher level of influence after the interview. All these accounts are verified and have a minimum of 0.4M 1 million followers (Supplement Table 3).

Most of the top influencers that appear in our dataset after the interview are prominent far-right media personalities with a large social media following. Many of the accounts that commented on and shared the interview possessed audience reach commensurate with mass media (DiResta, 2024). For example, the account **WarClandestine** (571K followers) propagated the false claim and QAnon conspiracy theory that Russia invaded Ukraine because the US had set up a network of biolabs there [23]. This account posted, "*Initial thoughts after the interview: -Putin* *knows way too much history -Putin is operating on an intellectual plane far above all US politicians -Putin appears to want cooperation, but the West has isolated Russia -Putin does not want to invade Poland or take over the world… https://t.co/VGiLYu74pD*" This was seen by 2.4 million viewers.

The **EndWokeness** account (3M followers) has been linked to false claims during the 2024 US presidential election regarding Haitian refugees eating people's pets in Springfield, Ohio [29]. This account posted *"Remember when the powers that be pushed Tucker out of Fox b/c he wasn't fully supportive of the WW3/NATO expansionist agenda? They thought they could silence him How did that work out for them?"* This post was seen by close to 5 million viewers according to the X's Impression metric.

Far-right influencers aligned with Tucker Carlson took it upon themselves to promote the interview and its content, especially after the interview. The only anti-Putin account whose comments had significant reach either before or after the interview was that of **Ron Filipkowski** (1M followers). He maintains sizable influence on social media, but his following and celebrity are a mere fraction of his right-wing counterparts.

### 4.2 Structure

After the interview, a more compact but less-defined network structure was formed compared to before. This indicates that the network structure potentially facilitated faster information flow while providing a weaker mechanism for checking information quality.

On the one hand, increased average degree and decreased network diameter after the interview (in Table 3) indicates a denser network with potentially stronger connectivity. A higher average degree indicates that nodes, on average, have more connections. Similarly, lower network diameter in the after-data indicates a more compact network formed after the interview (meaning one can reach to other nodes in shorter paths). In this environment, information can flow faster. On the other hand, a lower modularity value after the interview shows less defined community boundaries. The less modular structure of the after-data network (compared to the before-data) indicates that the interview may have made the network more vulnerable to disinformation spread. This is because weaker community boundaries reduce the network's ability to contain the information within separate clusters.

Table 3. Network structure before and after the interview

| | Avg. Degree | Network Diameter | Modularity |
|---|---|---|---|
| Before | 1.443 | 14 | 0.708* |
| After | 1.559 | 12 | 0.669* |
| | (p 8%) | (q 14%) | (q 5.5%) |

Note: This was run using the default resolution value of 1.0 from Gephi.

The visual analytics in Figure 4 show a shift in activity: right-wing individuals became more engaged after the interview (green community from 41% to 45 %), while left-wing interest declined (purple community: from 33% to 27%). This indicates that the right-wing community was energized by the interview while the left-wing community decreased its level of engagement.

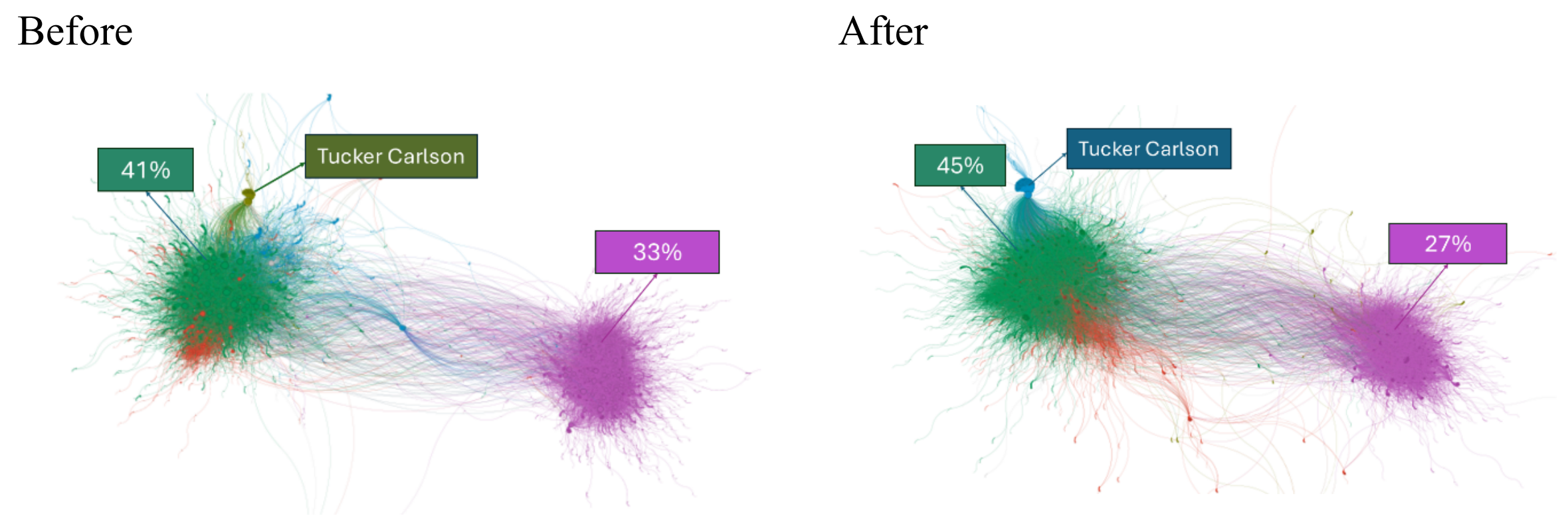

Figure 4. Visualization of before and after network structure and influencers. (Note: the percentage of community size is measured using the modularity algorithm in Gephi with resolution set to 2. Green represents the right-wing community and purple represents the left-wing community.)

### 4.3 Content: Topic Analysis

The UTs in Figure 5, using original posts, shows that the interview sharply intensified discussion over truth-telling as users grappled with the substance of Putin's assertions and debated their veracity. Discussion on the US and West increased mildly, while conversations about Putin and Russia rose considerably.

First, the UT of "Truth" includes keywords such as: interview, Putin, leader, journalists, Carlson, propaganda, truth, lies, lying, history, historical, believe, Biden, media and news. Users on X wondered who was telling the truth both before and after the interview (Appendix II includes topics assigned to each of the four UTs). The "Truth" UT experienced a sudden spike in activity, with around a 400% increase in volume after the interview.

During the interpretation, each topic in the "Truth" UT was labeled as either "Pro" or "Anti" Putin. Of the 10 topics, 7 supported Putin (48% of 13,769 posts), while 3 were critical (52% of 13,769 posts). Data is available in Appendix II. The posts supportive of Putin often defend his actions ("*I believe that Putin is not the total bad guy they make him out to be. Yes he has done bad things but he is trying to keep his country safe. He is not interested in taking over the world.*") or repeat Putin's narratives (i.e., "*In the war of propaganda, it's very difficult to defeat the United States. The United States controls all the world's media and many European media.").* The authors of these posts supported Putin's stance and believed he was truthful.

The value of triangulation is evident in our findings. Although the original posts reflected both pro- and anti- Putin narratives, it was the pro-Putin posts that gained the most traction. Synthesizing the UT results with our results concerning the audience and structure, it appears that discussions supporting Putin's narratives gained more attention and interaction from right-wing audiences.

Second, the UT of Putin and Russia attracted the attention of many users on X, as people criticized or supported Putin and the Russian position in the Russia-Ukrainian war (labeled as Putin_Russia in Figure 5). Some posts connect Putin's rationale with US domestic politics ("🇷🇺 *Putin has made it clear that he does not consider the United States 🇺🇸 or its people his enemy, but rather the corrupt ruling class that has taken over the United States. Putin is an enemy of the deep state. What if Russia is telling the truth? What if America/NATO is the bad…*").

Third, the UT of Ukraine and the war was about the ongoing debate over continued US military funding for Ukraine. 11 topics were assigned to this UT and the representative keywords included: *senate, money, funding, immigration, border, support, democracy, America, negotiations,* and *peace*. Users debated continued US funding for Ukraine and the Ukrainian President Volodymyr Zelensky's leadership (labeled as Ukraine_War in Figure 5). The interview occurred at a time when Congress was debating another round of funding for Ukraine. The $61B foreign aid package ultimately passed in April, but in the months leading up to the vote, the question of whether the US should continue to provide military assistance to Ukraine was widely debated on X. Some accounts expressed support for funding to Ukraine ("*And we do and will support Ukraine. Go with Tucker to Russia if you don't want to be an American. But in this country, we defend liberal democracy. Isolationism doesn't work. It never has.*"), while others objected to continued assistance (*"We want a good border bill, not money for Ukraine"*)

Fourth, another UT concerns criticism of the US, NATO, and other Western countries like Poland (this is labeled as US_West in Figure 5). The Kremlin has blamed the United States and what it describes as "the collective West" for the conflict, asserting that NATO expansion provoked Russia's invasion and that the war would end once US military assistance to Ukraine ends. Such a view, which largely discounts the agency and national self-determination of Ukrainians, is nonetheless shared by certain segments of the American right and left. Posts on X included, for example: *"Putin has no desire to expand the WAR in Poland or other NATO country.... the war talk coming from EU and USA is to get taxpayer dollars... If the USA stops funding Ukraine the war will end and negotiations will take place."*

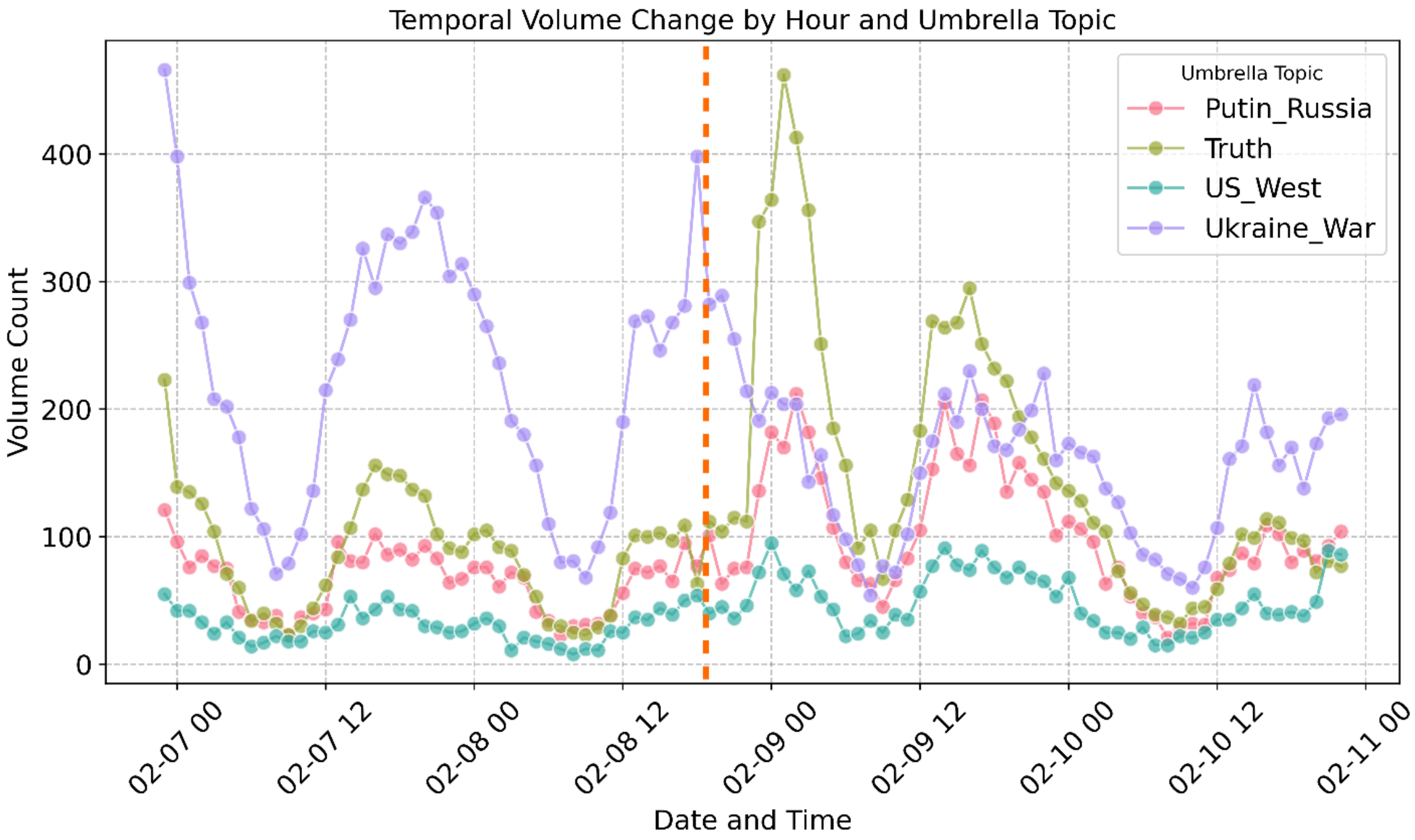

Figure 5. Temporal volume change of the umbrella topic (Note: The red dotted line indicates the interview release time: 6 PM EST, Feb 8th, 2024).

# 5 CONCLUSION

The present work seeks to develop a universal set of metrics for measuring the impact of a social media event. We proposed a framework to measure units of audience, structure, and content using ML and NLP techniques. Given the concern over relying on black-box algorithms and challenges of interpretation and validation of these tools, we used triangulation methods. The reported findings are drawn from evidence consistent across diverse data, method, and investigator triangulations. This study focused on a single case on the social media platform X. Therefore, the findings are limited to this platform with limited generalizability. Nonetheless, the framework of measuring social media's impact on an audience can be adapted to study other social media behaviors. Our triangulation approach is an initial attempt to incorporate a human in-the-loop of technical work. Further investigations are necessary to improve the rigor of the triangulation method applied to AI and ML approaches for social science inquiries.

Our analysis demonstrates that Putin's first major interview with Western media since Russia's full-scale invasion of Ukraine in February 2022 had a measurable effect on the X platform. The Russian president's interview revealed that: 1) the reach and engagement of far-right influencers increased after the interview, demonstrating the degree to which Kremlin narratives are shared among the far-right, 2) the communication structure became more vulnerable to disinformation spread after the interview, and 3) the public discourse reflected a change from support for Ukraine funding to conversations about Putin and Russia. In particular, the interview prompted many X users to grapple with the veracity of Putin's claims and to wonder where the truth lied.

We adopted triangulation methods (data, method, and investigator) to face methodological problems honestly without trying to avoid or downplay challenges. Our multidisciplinary approach derived from a desire to be vigilant and critical of our own methods and assumptions. From "the dilemmatic point of view" of McGrath (1982), "*all* research strategies and methods are seriously flawed" [26:179]. By drawing upon various methods and disciplines, our triangulation approach diminishes the overall impact of flaws inherent in any singular investigator or research method. In addition, the present work demonstrates that big data-driven methods, combined with rigorous validation through triangulation, provide compelling results that might not be revealed by other means. Our study showed that data-driven analysis can complement top-down and qualitative findings. In addition, a big data-driven approach enables finding hidden phenomena and provides actionable insights.

These findings contribute to research concerning the scope and nature of foreign adversarial influence. Vladimir Putin used his interview with Tucker Carlson to directly influence an American audience, with the goal of reducing US support for Ukraine. Our findings indicate that the interview had a broad and measurable impact on the X platform, especially among right-wing audiences. It is difficult to assess the long-term effect of this interview using social media data. However, given that many users spend considerable time on X and other platforms, they are susceptible to the "illusory truth effect" in which repeated exposure to certain claims increases belief in those claims (Lin et al., 2024). We do not assume that X posts have effects on behavioral changes. Rather, we investigated how this significant news event had a measurable impact on X communication patterns. The findings expose how "truth" is debated in social media after an injection of Putin's narratives. Instead of antagonizing the ideological stance of the opposing side, our findings highlighted which truth is in debate in crucial domain of foreign affairs. Social media's role in politics could be an unintended development. Nevertheless, it is our collective responsibility to foster a healthy and thriving social media ecosystem, rather than one filled with incivility and falsehood.

## Acknowledgments

This work was partially supported by grant from Cyber Florida.

# Supplement

## 1 Location validation

One author randomly sampled 50 posts with a location value. The X account profile was investigated to extract location information. If unavailable, the author analyzed post content to determine the user's location. The manual location annotation results matched the location information of the downloaded dataset in 47 out of 50 cases, resulting in a **94% agreement**. For three annotations, the annotator and Brandwatch disagreed. It appears Brandwatch inferred the location while the annotator was uncertain. For instance, a user listed their location as "Heaven," and Brandwatch tagged it as Indonesia. The other two cases were similar.

## 2 Tables

**Table 1. Descriptive statistics of the data**

| | | |
|---|---|---|
| Total # of posts | | 361,018 |
| | Original post | N= 44,889 |
| | Repost | N=145,821 |
| | Reply | N=61,739 |
| Original Posts including the reconstructed posts | | 190,710 |

**Table 2. Descriptive statistics of the corpora for BERT Topic Modeling (N=190,710)**

| | Average length of posts | Min length | Max length | Feq. Original posts included in the data |
|---|---|---|---|---|
| Before | 34 | 1 | 103 | 75,873 |
| After | 33 | 1 | 116 | 114,837 |

**Table 3. Number of followers and verification status of influencers before and after the interview**

| Account Name Before | # Followers | Verification Status | Account Name After | # Followers | Verification Status |
|---|---|---|---|---|---|
| TuckerCarlson | 14.7M (2024/11) | Verified | TuckerCarlson | 14.7M (2024/11) | Verified |
| WarClandestine | 0.6M (2024/11) | Verified | WarClandestine | 0.6M (2024/11) | Verified |
| simonateba | 0.6M (2024/11) | Verified | **EndWokeness** | 3.2M (2024/11) | Verified |
| bennyjohnson | 2.2M | Verified | **RonFilipkowski** | 1M (2024/11) | Verified |
| MattWallace888 | 1.5M | Verified | bennyjohnson | 3.1M (2024/11) | Verified |
| charliekirk11 | 4.0M | Verified | **CollinRugg** | 1.6M (2024/11) | Verified |
| VivekGRamaswamy | 3.2M (2024/11) | Verified | **VigilantFox** | 1.4M (2024/11) | Verified |
| seanmdav | 4.1M | Verified | **catturd2** | 3.1M (2024/11) | Verified |
| DavidSacks | 0.8M | Verified | **BasedMikeLee** | 0.4M (2024/11) | Verified |
| RealAlexJones | 2.1M | Verified | charliekirk11 | 4.1M (2024/11) | Verified |

Note: Some of the data did not record the number of followers during the study range. In that case, we used the most current number of followers in November 2024. Values are rounded to two decimal places, e.g., 571000 is 0.6M.

## 3 Figures

**Figure 1. The elbow method to determine the optimal clusters.**

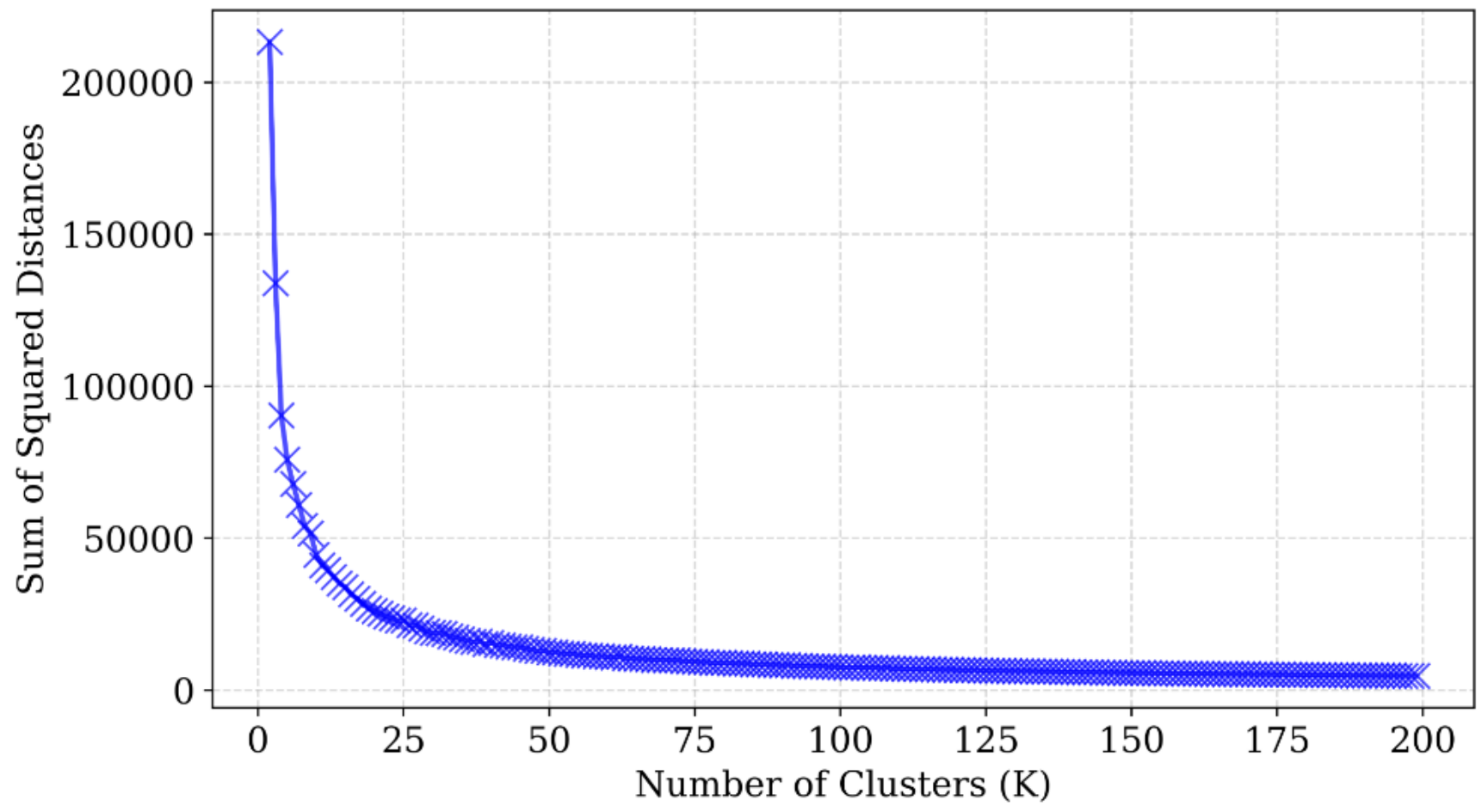

**Figure 2. Frequency of posts in the "before-" and "after-" datasets**

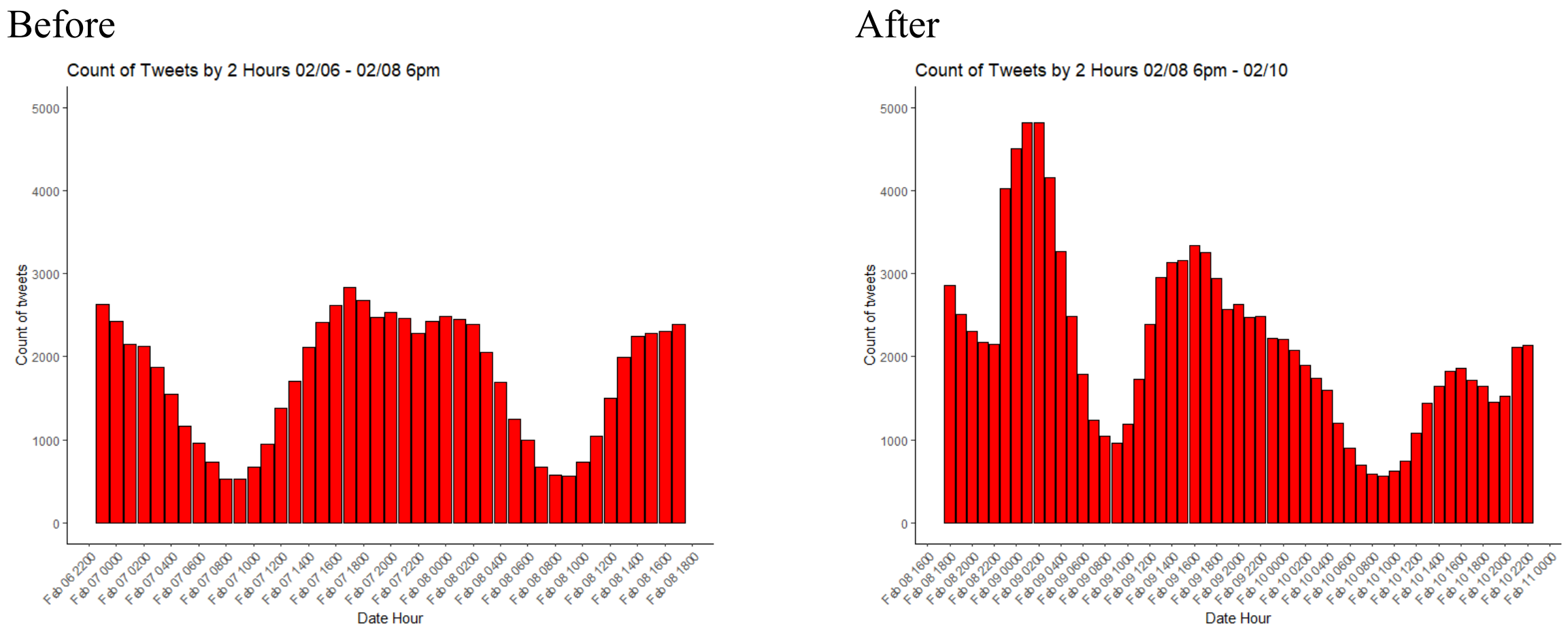

Note: The "before" and "after" dataset includes X posts created before and after 48 hours of the interview, respectively.

**Figure 3. Frequency of unique authors included in the “before” and “after” datasets**

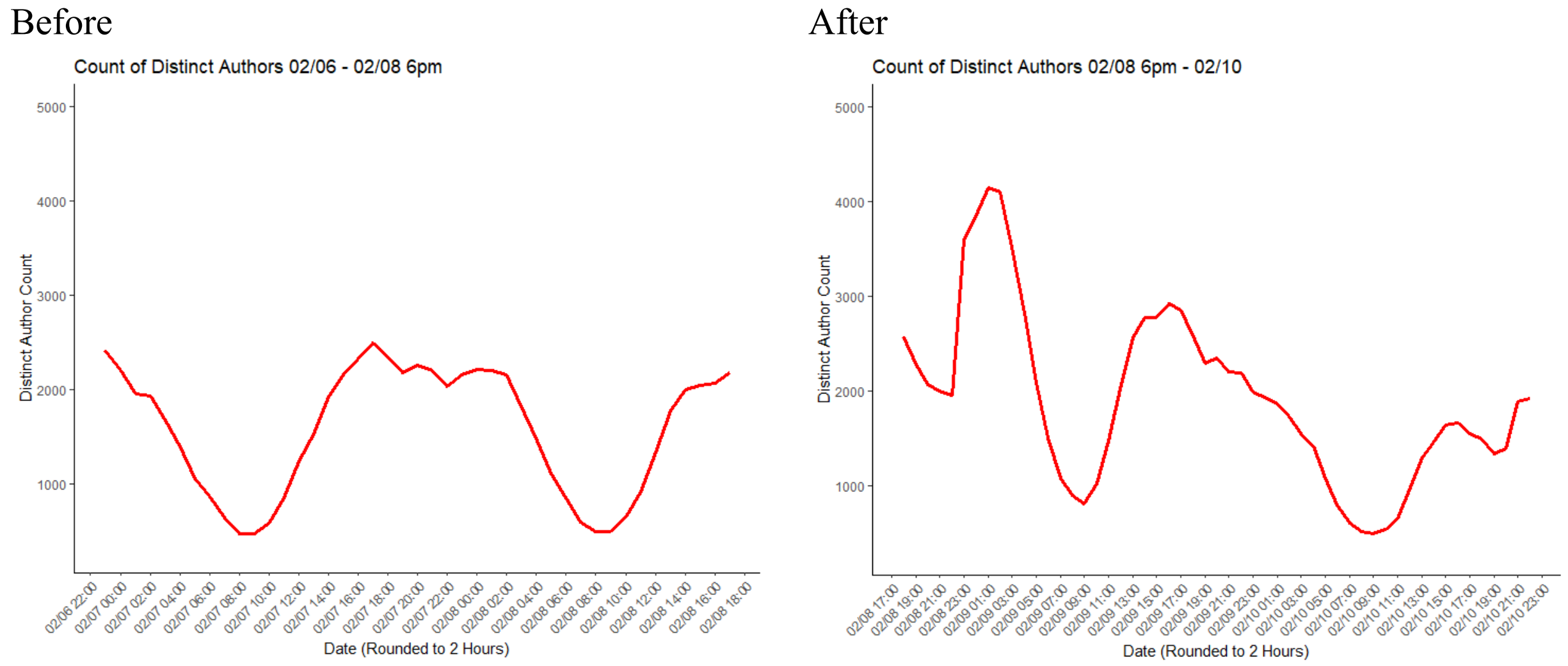

Note: The “before” dataset includes posts created 48 hours before the interview and the “after” dataset includes 48hours after the interview.

## Appendix I. 50 Topics and Representative Posts from BERT Topic Modeling

| Topic | Count | Name | Representation | Representative_Docs |
|---|---|---|---|---|
| 0 | 3133 | 0_putin_interview_vladimir_interviews | ['putin', 'interview', 'vladimir', 'interviews', 'russian', 'interviewing', 'interviewed', 'russia', 'journalist', 'journalists'] | ["If this guy has NOT watched the Putin interview, he and others in Government are not serious about anything. Seriously , if Putin is your enemy and you DON'T watch the interview. You're irresponsible and should not have the job you have.", "I didn't watch too many interviews by you, but probably this one is the worst interview you ever took. I didn't feel that I was watching western journalist interviewing Putin, but it looked more like Kremlin puppet journalists taking interview from Putin.", ' 💥 RUThe Vladimir Putin Interview'] |
| 1 | 2964 | 1_ukraine_dollars_funding_money | ['ukraine', 'dollars', 'funding', 'money', 'fund', 'pay', 'budget', 'aid', 'israel', 'war'] | ['NO MORE MONEY TO UKRAINE', 'No more money to Ukraine!!!', 'No more money for Ukraine!'] |
| 2 | 2725 | 2_border_borders_ukraine_bill | ['border', 'borders', 'ukraine', 'bill', 'immigration', 'funding', 'billion', 'fund', 'bills', 'illegals'] | ['We want a good border bill, not money for Ukraine', 'It's an Ukraine funding bill, not a border bill. When most of the money goes to Ukraine, calling it a border bill is a lie.', 'Did you read the bill? Why are we sending money to Ukraine or Israel if it's a border bill? It's not a border bill! So republicans won't sign 🤦 \u200d♀'] |
| 3 | 2208 | 3_putin_tucker_interviews_interview | ['putin', 'tucker', 'interviews', 'interview', 'interviewed', 'carlson', 'interviewing', 'journalists', 'journalist', 'cnn'] | ['Yeah well that's because most journalists fact check and Tucker doesn't, so of course Putin is going to have him be the one that does the interview. I'm sure Tucker will make him look good in his mind.', 'The Tucker interview with Vladimir Putin', 'Tucker's interview with President Putin is;'] |
| 4 | 2141 | 4_putin_vladimir_russia_trump | ['putin', 'vladimir', 'russia', 'trump', 'president', 'dick', 'him', 'with', 'clown', 'bitch'] | ['#Putin RU', '!!! #Putin', '#Putin'] |
| 5 | 1997 | 5_putin_kgb_vladimir_russia | ['putin', 'kgb', 'vladimir', 'russia', 'dictator', 'trump', 'evil', 'bad', 'president', 'leader'] | ['As bad as Putin is...there are things even he would not do to his country like USG is doing to us!', "Putin maybe bad but he has spoken so many truths and that's why the media hates him and puts him out to be way worse than he really is. The media has done the same to Trump so it is clear to see that whomever they call bad is actually the good one. This is why they are scared....", 'I believe that Putin is not the total bad guy they make him out to be. Yes he has done bad things but he is trying to keep his country safe. He is not interested in taking over the world.'] |
| 6 | 1945 | 6_nato_allies_alliance_russia | ['nato', 'allies', 'alliance', 'russia', 'putin', 'europe', 'eu', 'nations', 'ukraine', 'military'] | ['who signed the deal between Russia and NATO that NATO would not expand?', '"over NATO expansion" Lol that\'s such bullshit, Ukraine was not going to become part of NATO, and NATO expanded since the invasion', '"But NATO"'] |
| 7 | 1852 | 7_tucker_putin_russian_russia | ['tucker', 'putin', 'russian', 'russia', 'trump', 'cia', 'him', 'who', 'traitor', 'idiot'] | ['Tucker and Putin will.', 'Tucker in Russia.', 'Putin and tucker????? What???'] |
| 8 | 1840 | 8_ukrainewar_ukrainerussiawar_ukraine_missiles | ['ukrainewar', 'ukrainerussiawar', 'ukraine', 'missiles', 'kyiv', 'missile', 'ukrainian', 'attacked', 'russia', 'russians'] | ['russia is now attacking Ukraine with various missiles and drones Explosions are heard in Kharkiv. Mykolaiv is being attacked by drones, 🤬 there is a hit in a residential building, a house in Mykolaiv is burning...', 'Here's what we are reading today: Ukraine downed 44 of the 64 Russian missiles and drones that attacked Kyiv, Lviv, Mykolaiv, Dnipropetrovsk and Kharkiv regions on Wednesday morning. The attack took 4 lives and injured over 40 people.', '10:44 am in #Kyiv Morning from Ukraine! The night was calm in Kyiv. But the news from Kharkiv makes me sick and angry. russian drones attacked a gas station in Kharkiv at night, causing an oil spill that, in turn, set 14 private residential buildings on fire. A fire in a...'] |
| 9 | 1810 | 9_ukraine_ukrainians_ukrainian_russia | ['ukraine', 'ukrainians', 'ukrainian', 'russia', 'russians', 'russian', 'country', 'propaganda', 'nazis', 'ethnic'] | ['Ukraine=Russia', 'The is Ukraine', 'Ukraine is Russia...'] |
| 10 | 1797 | 10_putin_moscow_carlson_russian | ['putin', 'moscow', 'carlson', 'russian', 'tucker', 'vladimir', 'russia', 'interview', 'fox', 'interviews'] | ['The Putin/Carlson interview...', 'Tucker Carlson The Vladimir Putin Interview', 'Tucker Carlson: The Putin Interview'] |
| 11 | 1759 | 11_ukraine_russia_supporting_support | ['ukraine', 'russia', 'supporting', 'support', 'democracy', 'war', 'country', 'americans', 'america', 'right'] | ['The World can See that the , , and will not support Democracy at home, so we can only Assume this is why they want Ukraine to lose the War!!!!!!!!!!', "And we do and will support Ukraine. Go with Tucker to Russia if you don't want to be an American. But in this country we defend liberal democracy. Isolationism doesn't work. It never has", "If you don't support Ukraine, you support Putin."] |
| 12 | 1754 | 12_putin_propaganda_trump_cia | ['putin', 'propaganda', 'trump', 'cia', 'believe', 'truth', 'lies', 'lie', 'lying', 'says'] | ['You believe what Putin says?', 'The American people are literally 🤬 for transparency. Who knows how many people are going to believe everything that Putin says. Who knows what I'll believe. I know if the Government were truthful about Covid, just that nobody knew and we did the best we could. We did wrong,...', 'Putin has said every thing, & will say anything to promote himself as a God, dictator. Why does any one believe what he says? Of course, you believe what you want.'] |
| 13 | 1751 | 13_russia_putin_russian_ussr | ['russia', 'putin', 'russian', 'ussr', 'war', 'alaska', 'iran', 'china', 'korea', 'america'] | ['China will have non of it. China and Russia will fight over us.', '"RUSSIA RUSSIA RUSSIA" after all', 'The US is not at war with Russia.'] |
| 14 | 1677 | 14_carlson_putin_tucker_moscow | ['carlson', 'putin', 'tucker', 'moscow', 'russian', 'fox', 'russia', 'vladimir', 'boris', 'trump'] | ['Tucker Carlson and Putin be like:', 'Why is Tucker Carlson in Russia?', 'Putin to Carlson'] |
| 15 | 1635 | 15_putin_russia_russian_history | ['putin', 'russia', 'russian', 'history', 'historical', 'remember', 'president', 'lesson', 'speak', 'have'] | ['I thought I was the only one going on this amazing history lesson from #Putin .', 'Well Mr Woods I have see countless posts from you on the subject of history and historical events. I would love to hear from you on the history briefing on Russia that Valdimir Putin laid out . I was very intrigued by this. Thousands of years of history on the Russian people.', "Awesome history lesson by Putin. We, the West and especially America, have no idea how important history is to the Russians. We should not disregard what he's saying. It's important to understand why Russia thinks the way it does."] |
| 16 | 1619 | 16_ukraine_ukrainian_ukrainians_country | ['ukraine', 'ukrainian', 'ukrainians', 'country', 'us', 'america', 'we', 'our', 'support', 'fuck'] | ['Ukraine UA ❤ UA', 'F Ukraine and F you.', 'UA 🌻 👊 #Ukraine'] |
| 17 | 1604 | 17_ukraine_crimea_ukrainian_ukrainians | ['ukraine', 'crimea', 'ukrainian', 'ukrainians', 'putin', 'russia', 'war', 'russian', 'nato', 'nukes'] | ['Then, too, the war in Ukraine would never have started if Russia had not sent weapons of war to Ukraine', "You realize that the only way to prevent a nuclear war is to aid Ukraine, right? It's not a proxy war because we didn't start the war. We can't send our troops there to help because Putin threatened the US if we did. Because it'd be over already if we had put boots on the ground.", 'The war in Ukraine will not end with the defeat of Russia. Putin will not stop. For the war to end, Ukraine will have to give up some land. We can spend 10s of billions of dollars on a war that Ukraine cannot win, or we can push Ukraine to give up some land and end the war.'] |
| 18 | 1590 | 18_hitler_adolf_nazis_nazi | ['hitler', 'adolf', 'nazis', 'nazi', 'holocaust', 'mussolini', 'stalin', 'wwii', 'fascism', 'genocide'] | ['Trump is America's Hitler', 'So, Hitler, Hitler, and civil war that at this point Hitler would probably win.', 'Him and hitler.'] |
| 19 | 1589 | 19_putin_biden_vladimir_russia | ['putin', 'biden', 'vladimir', 'russia', 'trump', 'obama', 'president', 'believe', 'leader', 'better'] | ['Do you believe Putin over Biden?', 'Joe Biden is more evil than Vladimir Putin - Y / N 🇺🇸', 'How come Putin knows better than Biden? I think Biden knows nothing about both America and Russia. Is Biden unfit for being president?'] |
| 20 | 1541 | 20_senate_bipartisan_aid_senators | ['senate', 'bipartisan', 'aid', 'senators', 'senator', 'immigration', 'border', 'israel', 'funding', 'mcconnell'] | ['Senate advances aid bill for Ukraine, Israel and Taiwan without provisions for U.S. border', 'McConnell calls for the Senate to move on from the border security package - and still try to pass aid for Ukraine, Israel, and Taiwan.', 'US Senate blocks Ukraine, Israel aid bill  Senate Republicans voted against the bipartisan border security deal that was part of the $118 billion aid package, which included $60 billion for Ukraine, according to the Hill. Now, the Senate may take up the...'] |
| 21 | 1360 | 21_putin_tucker_tuckercarlson_interview | ['putin', 'tucker', 'tuckercarlson', 'interview', 'vladimir', 'interviews', 'carlson', 'fox', 'interviewing', 'watching'] | ['Tucker-Putin Interview', 'TUCKER CARLSON / PUTIN INTERVIEW ON X', 'The interview Tucker with Putin!'] |
| 22 | 1326 | 22_traitor_traitors_putin_treason | ['traitor', 'traitors', 'putin', 'treason', 'russia', 'russian', 'dictator', 'treasonous', 'patriot', 'moscowmike'] | ['YOU SURE DO LOVE YOUR DICTATORS! MOVE OUT OF OUR COUNTRY AND GO TO RUSSIA, TRAITOR', 'How is he a traitor? We are not at war with Russia.', 'If you support Putin you are a traitor'] |
| 23 | 1321 | 23_ukraine_putin_crimea_russia | ['ukraine', 'putin', 'crimea', 'russia', 'ukrainian', 'ukrainians', 'russians', 'rus', 'russian', 'invading'] | ['Long before the war in Ukraine? You're an idiot if you believe that. Putin invaded in 2014.', 'USA is the dictator and the Coup in Ukraine supported by the CIA is what started the war. The presidents of the USA have no power they wanted peace he said the CIA is the war machine they control it all. Ukraine was neutral and friendly with russia and russia was the same until...', "Putin said he was not going to invade Ukraine and he invaded it. Let's not say more."] |
| 24 | 1317 | 24_aid_ukraine_funding_assistance | ['aid', 'ukraine', 'funding', 'assistance', 'support', 'fund', 'war', 'give', 'democrats', 'maga'] | ['No. No more aid for Ukraine', "You want us to tell Congress? How about you tell Congress to support the Ukraine aid bill otherwise you'll pull funding. No aid to Ukraine means no aid for you.", 'No more aid to Ukraine'] |
| 25 | 1311 | 25_poland_polish_lithuania_ukraine | ['poland', 'polish', 'lithuania', 'ukraine', 'latvia', 'russia', 'germans', 'soviet', 'soviets', 'stalin'] | ['Poland 4.9% no way', 'So all Putin has to say to justify an invasion of Poland is that Poland forced him to do it.', 'People in Poland know.'] |
| 26 | 1303 | 26_putin_russia_trump_russian | ['putin', 'russia', 'trump', 'russian', 'dictator', 'war', 'president', 'enemy', 'deep', 'political'] | ['Trump wants America to be open for a political invasion by Putin. Best way to take over a nation is not through war, but through its politics. Why Putin is going to go down as one of the biggest dictators who tried to take over the world, but through the inside of government.', 'The only people who are against Putin are the people in congress. This is your war and not the people of America.', '🇷🇺 Putin has made it clear that he does not consider the United States 🇺🇸 or its people his enemy, but rather the corrupt ruling class that has taken over the United States. Putin is an enemy of the deep state. What if Russia is telling the truth? What if America/NATO is the bad...'] |
| 27 | 1245 | 27_sanctions_putin_tucker_nato | ['sanctions', 'putin', 'tucker', 'nato', 'carlson', 'russian', 'eu', 'russia', 'european', 'ukrainian'] | ['BREAKING: The European Union is said to be seeking sanctions and a "travel ban" against American journalist Tucker Carlson for his interview with Russian President Vladimir Putin. #ElonMusk says, If true, this would be disturbing indeed...One may agree with #TuckerCarlson or...', 'Tucker Carlson could face EU sanctions for interviewing Vladimir Putin A pair of European legislators are calling on the EU to place sanctions on Carlson following his interview of the Russian president.', '🚨 🚨 BREAKING: The European Union is said to be seeking sanctions and a "travel ban" against Tucker Carlson for his interview with Russian President Vladimir Putin.'] |
| 28 | 1201 | 28_ukraine_ukrainians_ukrainian_war | ['ukraine', 'ukrainians', 'ukrainian', 'war', 'kyiv', 'russia', 'russian', 'nato', 'allies', 'troops'] | ['Why the Ukraine War Is Not What You Think', 'Why we should care about it at all ? Main question is how this war is gonna end ? We know what Putin wants, we know what Ukraine wants, but problem is that Ukraine is not gonna win this war, US and EU failed with help . So what is next step ?', 'Things must be going worse than I thought in Ukraine if they are having to make women fight the war. Are all of the men dead? Time to negotiate peace and end this needless war that has killed so many Ukrainians. Americans will no longer be paying for Zelensky's war.'] |
| 29 | 1183 | 29_zelensky_zelenskyy_zaluzhny_zaluzhnyi | ['zelensky', 'zelenskyy', 'zaluzhny', 'zaluzhnyi', 'syrskyi', 'ukrainian', 'ukraine', 'syrsky', 'putin', 'russian'] | ['🇺🇦⚔🇷🇺 President Zelensky has promoted Colonel-General Oleksandr Syrsky to the position of the Commander-in-Chief of the Armed Forces of Ukraine, replacing Valery Zaluzhny.', " ⚡ Zelensky announces new chief  of general staff. Ukrainian President Volodymyr Zelensky has named Major General Anatoliy Barhilevych as the new Chief of General Staff for Ukraine's Armed Forces, Zelensky announced during his evening address on Feb. 9.", 'Now do Zelensky and Ukraine.'] |
| 30 | 1050 | 30_putin_maga_russian_russia | ['putin', 'maga', 'russian', 'russia', 'magas', 'republicans', 'gop', 'democrats', 'vladimir', 'republican'] | ['The Republicans are the party of Putin and this what they'd like us to believe. A vote for a Republican is a vote for Putin.', 'I hate what used to be the GOP now nothing more than the Trump/Putin Nazi MAGA party', 'The Putin Republican MAGA Party'] |
| 31 | 973 | 31_collusion_putin_mueller_conspiracy | ['collusion', 'putin', 'mueller', 'conspiracy', 'russian', 'colluded', 'russia', 'fbi', 'hoax', 'investigation'] | ['He was the agent in charge of the investigation into the trump Russia collusion thing, he was found to be a double agent, and you don't see how that would've helped trump? trump owes Russians and putin money, that's how he's been getting financed, bc he's gas 7 bankruptcies', "You mean like Trump/Russia a proven Hoax you're right but the only person who colluded with a Russian was Hillary Clinton", 'The documents case has always been BS. Trump had the right to the documents. Calling him a security threat was a repeat of the Russia collusion hoax. In other words, BS.'] |
| 32 | 956 | 32_tuckercarlson_putin_putininterview_tuckerputin | ['tuckercarlson', 'putin', 'putininterview', 'tuckerputin', 'tucker', 'tuckerputininterview', 'vladimir', 'vladimirputin', 'russian', 'carlson'] | ['👀 #TuckerCarlson #putin', 'The Vladimir Putin Interview..  #TuckerCarlson', 'Tucker - Putin interview #TuckerCarlson'] |
| 33 | 926 | 33_ukraine_crimea_boris_putin | ['ukraine', 'crimea', 'boris', 'putin', 'ukrainian', 'negotiations', 'negotiate', 'peace', 'negotiated', 'talks'] | ['President Putin has said he wants to negotiate peace several times! ukraine agreed to terms in Istanbul and backed out thanks to boris johnson and I'm sure our corrupt government. This could have been over long ago.', 'That Peace Deal that was supposed to happened 18 months ago with Russia and Ukraine. You can blame Boris Johnson for that for the interference. I think that's why they are out there because of Putin said in the interview.', 'You must listen to the 2 hour interview to pick up on things I have already been telling you about Ukraine. There have been attempted peace negotiations as I said.Boris Johnson went in & told Zelensky not to listen to Putin on peace that the west is not ready for the war to end.'] |
| 34 | 916 | 34_putin_propaganda_media_russia | ['putin', 'propaganda', 'media', 'russia', 'journalism', 'journalists', 'journalist', 'russian', 'news', 'cia'] | ['Putin says Western media is just propaganda to benefit "American financial institutions": "In the war of propaganda, it\'s very difficult to defeat the United States. The United States controls all the world\'s media and many European media."', 'CITIZEN FREE PRESS Putin says Western media is just propaganda to benefit "American financial institutions": "In the war of propaganda, it\'s very difficult to defeat the United States. The United States controls all the world\'s media and many European media."', "Putin to Tucker on the U.S. role in media: "In the war of propaganda, it is very difficult to defeat the United States because the United States controls all the world's media and many European media.""] |
| 35 | 869 | 35_putin_russia_nato_ukraine | ['putin', 'russia', 'nato', 'ukraine', 'syria', 'russian', 'war', 'iran', 'nukes', 'europe'] | ['Putin has no desire to expand the WAR in Poland or other NATO country.... the war talk coming from EU and USA is to get taxpayer dollars... If the USA stops funding Ukraine the war will end and negotiations will take place.', 'Putin is saying what many of us have believed from the start. To be clear, I am not a Putin nor a Zelensky fan. And even though I do not agree with the war, I can also still understand why Russia has chosen to go to war. The expansion of NATO countries, the violations of Minsk...', "Russia has been trying to expand for several years now. We can't go back to the Cold War years. We need to finish what we started. By not doing so is now they (and we) lose. Putin doesn't want peace. He never has."] |
| 36 | 858 | 36_russia_putin_russian_enemy | ['russia', 'putin', 'russian', 'enemy', 'america', 'communist', 'republicans', 'trump', 'china', 'country'] | ['Russia is not your enemy.', 'Pro-Russia America First crowd? So which one is it? Pro-Russia? Or America First? Or is Russia the new America?', 'You support Russia over America. Why are you in this country?'] |
| 37 | 835 | 37_biden_ukraine_bidens_trump | ['biden', 'ukraine', 'bidens', 'trump', 'vp', 'obama', 'burisma', 'war', 'documents', 'president'] | ['Biden has done more for Ukraine', 'Trump and Biden are for gun control Trump and Biden spend without any accountability Trump and Biden had horrid Covid policy that destroyed lives. Trump and Biden are pro choice. Trump and Biden have funded Ukraine. Trump and Biden are against fighting a culture war and...', "I have nothing against Putin over the Ukraine, Biden is the cause of their war and Biden is the reason it's still going on. And when I say Biden make no mistake about it, I mean Obama"] |
| 38 | 780 | 38_russia_moscow_move_moving | ['russia', 'moscow', 'move', 'moving', 'stay', 'america', 'country', 'live', 'leave', 'comrade'] | ['MOVE TO RUSSIA!', 'Move. To. Russia. 🙄', 'Move to Russia RU'] |
| 39 | 635 | 39_currency_dollar_usd_putin | ['currency', 'dollar', 'usd', 'putin', 'sanctions', 'gdp', 'yuan', 'inflation', 'trillion', 'oil'] | ['BREAKING: President Putin says the US Government is killing the dollar with their own hands, many countries including major oil producers, are accepting payments for oil in the Chinese Yuan.', 'He is 100% correct on this. Our government is killing the dollar all by itself. Vladimir Putin says the US Government is killing the dollar with their own hands, many countries including major oil producers, are accepting payments for oil in Chinese yuan.', 'Vladimir Putin says the US Government is killing the dollar with their own hands, many countries including major oil producers, are accepting payments for oil in Chinese yuan.. not the U.S dollar. "They're $33 trillion in debt & they won't stop printing money"'] |
| 40 | 632 | 40_putin_russian_trump_vladimir | ['putin', 'russian', 'trump', 'vladimir', 'russia', 'biden', 'gop', 'tuberville', 'maga', 'president'] | ['Tommy Tuberville thinks if he stops the US from giving aid to Ukraine, Putin will stop invading Ukraine. I would usually have a difficult time thinking a United States Senator could possibly be this stupid, but it is Tommy Tuberville.', 'Mike Johnson, Who are You working for ? 1. Americans who Pay You ? OR 2. Comrade Putin & Trump ?', "So you're an idiot and a traitor willing to sell the nation out to our enemies to win the next election. Got it, Tuberville. How much did Putin pay you?"] |
| 41 | 617 | 41_orthodox_putin_bolshevik_russian | ['orthodox', 'putin', 'bolshevik', 'russian', 'russians', 'russia', 'ukrainian', 'ukraine', 'bolsheviks', 'soviet'] | ['"God of war': Russian Orthodox Church stands by Putin, but at what cost?', 'Surely there are people who identify as Rus and Orthodox Christian who feel a loyalty to their nation and are fighting for Putin. Conversely, there are some who identify more as Ukrainian in the mold of Bandera and may or may not be Christian but they wish to have their own land', 'No one is talking about how Putin said in the interview that he believes in God and wants to allow the people of Ukraine to be able to return to the Russian Orthodox Church.'] |
| 42 | 601 | 42_satellite_satellites_gps_starlink | ['satellite', 'satellites', 'gps', 'starlink', 'ukraine', 'russia', 'spacex', 'starlinks', 'russians', 'russian'] | ['This is taken from a satellite.', 'Russia is using SpaceX's Starlink satellite devices in Ukraine, sources say', 'Russia is using SpaceX's Starlink satellite devices in Ukraine \|'] |
| 43 | 588 | 43_israel_ukraine_isreal_hamas | ['israel', 'ukraine', 'isreal', 'hamas', 'israeli', 'palestine', 'gaza', 'palestinians', 'palestinian', 'iran'] | ['People keep trying to compare the crisis in Gaza with the Ukraine conflict. Let me help. Israel is to Gaza as Ukraine was to the Donbas. Russia is to Ukraine like the Arab world would be to Israel if any of the Arab leaders believed in Palestine like Putin believes in Russia.', 'In Ukraine or Israel? War? Why not?', "BE Careful, PUTIN is going to attack Israel. Biden & Blinken are openly opposing Israel's need to remove Hamas terrorist. PLEASE ASK the American people to donate to Israel so that America can be blessed rather than cursed. SUPPORT ISRAEL NO CEASE-FIRE UNTIL NO HAMAS!"] |
| 44 | 521 | 44_putin_pelosi_russian_hilary | ['putin', 'pelosi', 'russian', 'hilary', 'russia', 'hillary', 'clinton', 'her', 'uranium', 'she'] | ["Stefanik isn't bashful, About admitting she's a fascist, She spends all of her time, Kissing Trump and Putin's asses, She knows once her terms up, She'll never be in congress again, So, she helps the GOP, Replace democracy with fascism.", 'Isn't she in Russia.', 'Hillary sold uranium to Russia..... She is bought and paid for by Russia.'] |
| 45 | 445 | 45_wwii_ww2_wwi_1945 | ['wwii', 'ww2', 'wwi', '1945', '1944', '1941', 'war', 'historicalfiction', 'history', 'hitler'] | ['WWII started 3 years before that.', 'Was WWII ?', 'In wwii? Why using x?'] |
| 46 | 288 | 46_pleasure_bridge_indulge_read | ['pleasure', 'bridge', 'indulge', 'read', 'here', 'love', 'link', 'free', 'on', 'please'] | ['Send me a message via the link in the comments. Delight in my free-of-charge Bridge-of-love page for your pleasure! 📲 💕 . .', 'I crave to awaken the fire within you, fueling your most profound jones ! 🔥 Click on the link in the commentary. Gain free access to my Bridge-of-love and indulge yourself in measureless pleasure! 😘 💥 . .', 'My desire is to awaken the passion within you, fulfilling your most profound letch! 🔥 Do not miss out on clicking the link in the exposition. Access my Bridge-of-love for free and indulge yourself in pure bliss! 😌 🆓 . .'] |
| 47 | 222 | 47_putin_russian_vladimir_cia | ['putin', 'russian', 'vladimir', 'cia', 'russia', 'nordstream', 'tucker', 'nord', 'tuckercarlson', 'pipeline'] | ['RUUS ⚡ Tucker Carlson asks President Putin who blew up the Nord Stream: Tucker: "Who blew up Nord Stream?" Putin: "You for sure." Tucker: "I was busy that day. I did not blow up Nord Stream. Thank you though." Putin: "You personally may have an alibi, but the CIA has no such...', ' 🚨 🚨 NORD STREAM PIPELINE : "RUSSIA CLAIMS USA/CIA DID IT" 🚨 TUCKER TO PUTIN: "Who blew up the Nord Stream pipeline?" PUTIN: "You, of course." TUCKER: "I was busy that day. I did not blow up Nord Stream. Thank you though." PUTIN: "You may personally have an alibi, but the...', 'Putin fingers the CIA as the culprit who blew up Nord Stream Pipeline. Tucker: "Who blew up Nord Stream?" Putin: "You for sure." Tucker: "I was busy that day. I did not blow up Nord Stream." Putin: "You personally may have an alibi, but the CIA has no such." Do you...'] |
| 48 | 162 | 48_musk_putin_trump_russia | ['musk', 'putin', 'trump', 'russia', 'mullah', 'elon', 'smart', 'shapiro', 'tesla', 'him'] | ['Putin on Elon Musk: "I think Elon Musk is a smart person, I truly believe he is. There\'s no stopping Elon Musk"', 'Putin on Elon Musk:', 'Putin on Elon Musk “There’s no stopping Elon Musk”'] |
| 49 | 153 | 49_putin_trudeau_ukrainian_russian | ['putin', 'trudeau', 'ukrainian', 'russian', 'ukraine', 'zelensky', 'canadians', 'canadian', 'canada', 'nazi'] | ['Putin talks about Trudeau giving a standing ovation to an actual Waffen SS Nazi in Canadian Parliament "The President of Ukraine visited Canada. This story is well known but being silenced in the Western countries… The President of Ukraine stood up with the entire parliament…', '❗ US 🎙 RU - Tucker Carlson Interview: Putin accused Zelensky and Trudeau of giving a Nazi a standing ovation in the Canadian parliament: "The President of Ukraine visited Canada. This story is well known but being silenced in the Western countries… The President of Ukraine…', 'Putin brought up how Zelensky and Trudeau gave a standing ovation to a Nazi in the Canadian Parliament and says that Western Media has been censoring this. The President of Ukraine stood up with the entire parliament of Canada and applauded this man (Nazi). How can this be?💊 🔥 '] |

# Appendix II. Umbrella Topics and the Assigned Topics

| Umberella Topic | Topic ID | Count | Topic Representation | Representative_Docs | Stance to Putin |
|---|---|---|---|---|---|
| Truth | 0 | 3133 | ['putin', 'interview', 'vladimir', 'interviews', 'russian', 'interviewing', 'interviewed', 'russia', 'journalist', 'journalists'] | ["If this guy has NOT watched the Putin interview, he and others in Government are not serious about anything. Seriously , if Putin is your enemy and you DON'T watch the interview. You're irresponsible and should not have the job you have.", "I didn't watch too many interviews by you, but probably this one is the worst interview you ever took. I didn't feel that I was watching western journalist interviewing Putin, but it looked more like Kremlin puppet journalists taking interview from Putin.", '💥RUThe Vladimir Putin Interview'] | Anti |
| Truth | 3 | 2208 | ['putin', 'tucker', 'interviews', 'interview', 'interviewed', 'carlson', 'interviewing', 'journalists', 'journalist', 'cnn'] | ['Yeah well that's because most journalists fact check and Tucker doesn't, so of course Putin is going to have him be the one that does the interview. I'm sure Tucker will make him look good in his mind.', 'The Tucker interview with Vladimir Putin', 'Tucker's interview with President Putin is;'] | Anti |
| Truth | 5 | 1997 | ['putin', 'kgb', 'vladimir', 'russia', 'dictator', 'trump', 'evil', 'bad', 'president', 'leader'] | ['As bad as Putin is…there are things even he would not do to his country like USG is doing to us!', "Putin maybe bad but he has spoken so many truths and that's why the media hates him and puts him out to be way worse than he really is. The media has done the same to Trump so it is clear to see that whomever they call bad is actually the good one. This is why they are scared.…", 'I believe that Putin is not the total bad guy they make him out to be. Yes he has done bad things but he is trying to keep his country safe. He is not interested in taking over the world.'] | Pro |
| Truth | 12 | 1754 | ['putin', 'propaganda', 'trump', 'cia', 'believe', 'truth', 'lies', 'lie', 'lying', 'says'] | ['You believe what Putin says?', 'The American people are literally 🤐 for transparency. Who knows how many people are going to believe everything that Putin says. Who knows what I'll believe. I know if the Government were truthful about Covid, just that nobody knew and we did the best we could. We did wrong,…', 'Putin has said every thing, & will say anything to promote himself as a God, dictator. Why does any one believe what he says? Of course, you believe what you want.'] | Anti |
| Truth | 15 | 1635 | ['putin', 'russia', 'russian', 'history', 'historical', 'remember', 'president', 'lesson', 'speak', 'have'] | ['I thought I was the only one going on this amazing history lesson from #Putin .', 'Well Mr Woods I have see countless posts from you on the subject of history and historical events. I would love to hear from you on the history briefing on Russia that Valdimir Putin laid out . I was very intrigued by this. Thousands of years of history on the Russian people.', "Awesome history lesson by Putin. We, the West and especially America, have no idea how important history is to the Russians. We should not disregard what he's saying. It's important to understand why Russia thinks the way it does."] | Pro |
| Truth | 19 | 1589 | ['putin', 'biden', 'vladimir', 'russia', 'trump', 'obama', 'president', 'believe', 'leader', 'better'] | ['Do you believe Putin over Biden?', 'Joe Biden is more evil than Vladimir Putin - Y / N US', 'How come Putin knows better than Biden? I think Biden knows nothing about both America and Russia. Is Biden unfit for being president?'] | Pro |
| Truth | 34 | 916 | ['putin', 'propaganda', 'media', 'russia', 'journalism', 'journalists', 'journalist', 'russian', 'news', 'cia'] | ['Putin says Western media is just propaganda to benefit "American financial institutions": "In the war of propaganda, it\\'s very difficult to defeat the United States. The United States controls all the world\\'s media and many European media."', 'CITIZEN FREE PRESS Putin says Western media is just propaganda to benefit "American financial institutions": "In the war of propaganda, it\\'s very difficult to defeat the United States. The United States controls all the world\\'s media and many European media."', "Putin to Tucker on the U.S. role in media: "In the war of propaganda, it is very difficult to defeat the United States because the United States controls all the world's media and many European media.""] | Pro |
| Truth | 47 | 222 | ['putin', 'russian', 'vladimir', 'cia', 'russia', 'nordstream', 'tucker', 'nord', 'tuckercarlson', 'pipeline'] | ['RUUS ⚡ Tucker Carlson asks President Putin who blew up the Nord Stream: Tucker: "Who blew up Nord Stream?" Putin: "You for sure." Tucker: "I was busy that day. I did not blow up Nord Stream. Thank you though." Putin: "You personally may have an alibi, but the CIA has no such…', '🚨 🚨 NORD STREAM PIPELINE : "RUSSIA CLAIMS USA/CIA DID IT" 🚨 TUCKER TO PUTIN: "Who blew up the Nord Stream pipeline?" PUTIN: "You, of course." TUCKER: "I was busy that day. I did not blow up Nord Stream. Thank you though." PUTIN: "You may personally have an alibi, but the…', 'Putin fingers the CIA as the culprit who blew up Nord Stream Pipeline. Tucker: "Who blew up Nord Stream?" Putin: "You for sure." Tucker: "I was busy that day. I did not blow up Nord Stream." Putin: "You personally may have an alibi, but the CIA has no such." Do you…'] | Pro |
| Truth | 48 | 162 | ['musk', 'putin', 'trump', 'russia', 'mullah', 'elon', 'smart', 'shapiro', 'tesla', 'him'] | ['Putin on Elon Musk: "I think Elon Musk is a smart person, I truly believe he is. There\\'s no stopping Elon Musk"', 'Putin on Elon Musk:', 'Putin on Elon Musk "There's no stopping Elon Musk""] | Pro |
| Truth | 49 | 153 | ['putin', 'trudeau', 'ukrainian', 'russian', 'ukraine', 'zelensky', 'canadians', 'canadian', 'canada', 'nazi'] | ['Putin talks about Trudeau giving a standing ovation to an actual Waffen SS Nazi in Canadian Parliament "The President of Ukraine visited Canada. This story is well known but being silenced in the Western countries... The President of Ukraine stood up with the entire parliament…', '❗ US 🎙 RU - Tucker Carlson Interview: Putin accused Zelensky and Trudeau of giving a Nazi a standing ovation in the Canadian parliament: "The President of Ukraine visited Canada. This story is well known but being silenced in the Western countries... The President of Ukraine…', 'Putin brought up how Zelensky and Trudeau gave a standing ovation to a Nazi in the Canadian Parliament and says that Western Media has been censoring this. The President of Ukraine stood up with the entire parliament of Canada and applauded this man (Nazi). How can this be?💊 💧 '] | Pro |
| Total # of documents (Truth) | | 13769 | | Anti Putin: 52%, Pro Putin: 48% | |
| Putin_Russia | 5 | 1997 | ['putin', 'kgb', 'vladimir', 'russia', 'dictator', 'trump', 'evil', 'bad', 'president', 'leader'] | ['As bad as Putin is…there are things even he would not do to his country like USG is doing to us!', "Putin maybe bad but he has spoken so many truths and that's why the media hates him and puts him out to be way worse than he really is. The media has done the same to Trump so it is clear to see that whomever they call bad is actually the good one. This is why they are scared.…", 'I believe that Putin is not the total bad guy they make him out to be. Yes he has done bad things but he is trying to keep his country safe. He is not interested in taking over the world.'] | |
| Putin_Russia | 22 | 1326 | ['traitor', 'traitors', 'putin', 'treason', 'russia', 'russian', 'dictator', 'treasonous', 'patriot', 'moscowmike'] | ['YOU SURE DO LOVE YOUR DICTATORS! MOVE OUT OF OUR COUNTRY AND GO TO RUSSIA, TRAITOR', 'How is he a traitor? We are not at war with Russia.', 'If you support Putin you are a traitor'] | |
| Putin_Russia | 23 | 1321 | ['ukraine', 'putin', 'crimea', 'russia', 'ukrainian', 'ukrainians', 'russians', 'rus', 'russian', 'invading'] | ['Long before the war in Ukraine? You're an idiot if you believe that. Putin invaded in 2014.', 'USA is the dictator and the Coup in Ukraine supported by the CIA is what started the war. The presidents of the USA have no power they wanted peace he said the CIA is the war machine they control it all. Ukraine was neutral and friendly with russia and russia was the same until…', "Putin said he was not going to invade Ukraine and he invaded it. Let's not say more."] | |
| Putin_Russia | 26 | 1303 | ['putin', 'russia', 'trump', 'russian', 'dictator', 'war', 'president', 'enemy', 'deep', 'political'] | ['Trump wants America to be open for a political invasion by Putin. Best way to take over a nation is not through war, but through its politics. Why Putin is going to go down as one of the biggest dictators who tried to take over the world, but through the inside of government.', 'The only people who are against Putin are the people in congress. This is your war and not the people of America.', 'RU Putin has made it clear that he does not consider the United States US or its people his enemy, but rather the corrupt ruling class that has taken over the United States. Putin is an enemy of the deep state. What if Russia is telling the truth? What if America/NATO is the bad…'] | |
| Putin_Russia | 30 | 1050 | ['putin', 'maga', 'russian', 'russia', 'magas', 'republicans', 'gop', 'democrats', 'vladimir', 'republican'] | ['The Republicans are the party of Putin and this what they'd like us to believe. A vote for a Republican is a vote for Putin.', 'I hate what used to be the GOP now nothing more than the Trump/Putin Nazi MAGA party', 'The Putin Republican MAGA Party'] | |
| Putin_Russia | 35 | 869 | ['putin', 'russia', 'nato', 'ukraine', 'syria', 'russian', 'war', 'iran', 'nukes', 'europe'] | ['Putin has no desire to expand the WAR in Poland or other NATO country.... the war talk coming from EU and USA is to get taxpayer dollars... If the USA stops funding Ukraine the war will end and negotiations will take place.', 'Putin is saying what many of us have believed from the start. To be clear, I am not a Putin nor a Zelensky fan. And even though I do not agree with the war, I can also still understand why Russia has chosen to go to war. The expansion of NATO countries, the violations of Minsk…', "Russia has been trying to expand for several years now. We can't go back to the Cold War years. We need to finish what we started. By not doing so is now they (and we) lose. Putin doesn't want peace. He never has."] | |
| Putin_Russia | 36 | 858 | ['russia', 'putin', 'russian', 'enemy', 'america', 'communist', 'republicans', 'trump', 'china', 'country'] | ['Russia is not your enemy.', 'Pro-Russia America First crowd? So which one is it? Pro-Russia? Or America First? Or is Russia the new America?', 'You support Russia over America. Why are you in this country?'] | |
| Putin_Russia | 38 | 780 | ['russia', 'moscow', 'move', 'moving', 'stay', 'america', 'country', 'live', 'leave', 'comrade'] | ['MOVE TO RUSSIA!', 'Move. To. Russia. 🙄', 'Move to Russia RU'] | |

| | | | | |
|---|---|---|---|---|
| Putin_Russia | 41 | 617 | ['orthodox', 'putin', 'bolshevik', 'russian', 'russians', 'russia', 'ukrainian', 'ukraine', 'bolsheviks', 'soviet'] | ['‘God of war’: Russian Orthodox Church stands by Putin, but at what cost?', 'Surely there are people who identify as Rus and Orthodox Christian who feel a loyalty to their nation and are fighting for Putin. Conversely, there are some who identify more as Ukrainian in the mold of Bandera and may or may not be Christian but they wish to have their own land', 'No one is talking about how Putin said in the interview that he believes in God and wants to allow the people of Ukraine to be able to return to the Russian Orthodox Church.'] |
| US_West | 23 | 1321 | ['ukraine', 'putin', 'crimea', 'russia', 'ukrainian', 'ukrainians', 'russians', 'rus', 'russian', 'invading'] | ['Long before the war in Ukraine? You’re an idiot if you believe that. Putin invaded in 2014.', 'USA is the dictator and the Coup in Ukraine supported by the CIA is what started the war. The presidents of the USA have no power they wanted peace he said the CIA is the war machine they control it all. Ukraine was neutral and friendly with russia and russia was the same until…', "Putin said he was not going to invade Ukraine and he invaded it. Let's not say more."] |
| US_West | 30 | 1050 | ['putin', 'maga', 'russian', 'russia', 'magas', 'republicans', 'gop', 'democrats', 'vladimir', 'republican'] | ['The Republicans are the party of Putin and this what they’d like us to believe. A vote for a Republican is a vote for Putin.', 'I hate what used to be the GOP now nothing more than the Trump/Putin Nazi MAGA party', 'The Putin Republican MAGA Party'] |
| US_West | 35 | 869 | ['putin', 'russia', 'nato', 'ukraine', 'syria', 'russian', 'war', 'iran', 'nukes', 'europe'] | ['Putin has no desire to expand the WAR in Poland or other NATO country.... the war talk coming from EU and USA is to get taxpayer dollars... If the USA stops funding Ukraine the war will end and negotiations will take place.', 'Putin is saying what many of us have believed from the start. To be clear, I am not a Putin nor a Zelensky fan. And even though I do not agree with the war, I can also still understand why Russia has chosen to go to war. The expansion of NATO countries, the violations of Minsk…', "Russia has been trying to expand for several years now. We can't go back to the Cold War years. We need to finish what we started. By not doing so is now they (and we) lose. Putin doesn't want peace. He never has."] |
| US_West | 6 | 1945 | ['nato', 'allies', 'alliance', 'russia', 'putin', 'europe', 'eu', 'nations', 'ukraine', 'military'] | ['who signed the deal between Russia and NATO that NATO would not expand?', '"over NATO expansion" Lol that\'s such bullshit, Ukraine was not going to become part of NATO, and NATO expanded since the invasion', '“But NATO”'] |
| US_West | 25 | 1311 | ['poland', 'polish', 'lithuania', 'ukraine', 'latvia', 'russia', 'germans', 'soviet', 'soviets', 'stalin'] | ['Poland 4.9% no way', 'So all Putin has to say to justify an invasion of Poland is that Poland forced him to do it.', 'People in Poland know.'] |
| US_West | 39 | 635 | ['currency', 'dollar', 'usd', 'putin', 'sanctions', 'gdp', 'yuan', 'inflation', 'trillion', 'oil'] | ['BREAKING: President Putin says the US Government is killing the dollar with their own hands, many countries including major oil producers, are accepting payments for oil in the Chinese Yuan.', 'He is 100% correct on this. Our government is killing the dollar all by itself. Vladimir Putin says the US Government is killing the dollar with their own hands, many countries including major oil producers, are accepting payments for oil in Chinese yuan.', 'Vladimir Putin says the US Government is killing the dollar with their own hands, many countries including major oil producers, are accepting payments for oil in Chinese yuan.. not the U.S dollar. “They’re $33 trillion in debt & they won’t stop printing money”'] |
| Ukraine_War | 23 | 1321 | ['ukraine', 'putin', 'crimea', 'russia', 'ukrainian', 'ukrainians', 'russians', 'rus', 'russian', 'invading'] | ['Long before the war in Ukraine? You’re an idiot if you believe that. Putin invaded in 2014.', 'USA is the dictator and the Coup in Ukraine supported by the CIA is what started the war. The presidents of the USA have no power they wanted peace he said the CIA is the war machine they control it all. Ukraine was neutral and friendly with russia and russia was the same until…', "Putin said he was not going to invade Ukraine and he invaded it. Let's not say more."] |
| Ukraine_War | 1 | 2964 | ['ukraine', 'dollars', 'funding', 'money', 'fund', 'pay', 'budget', 'aid', 'israel', 'war'] | ['NO MORE MONEY TO UKRAINE', 'No more money to Ukraine!!!', 'No more money for Ukraine!'] |
| Ukraine_War | 2 | 2725 | ['border', 'borders', 'ukraine', 'bill', 'immigration', 'funding', 'billion', 'fund', 'bills', 'illegals'] | ['We want a good border bill, not money for Ukraine', 'It’s an Ukraine funding bill, not a border bill. When most of the money goes to Ukraine, calling it a border bill is a lie.', 'Did you read the bill? Why are we sending money to Ukraine or Israel if it’s a border bill? It’s not a border bill! So republicans won’t sign 🤦 \u200d♀'] |
| Ukraine_War | 8 | 1840 | ['ukrainewar', 'ukrainerussiawar', 'ukraine', 'missiles', 'kyiv', 'missile', 'ukrainian', 'attacked', 'russia', 'russians'] | ['russia is now attacking Ukraine with various missiles and drones Explosions are heard in Kharkiv. Mykolaiv is being attacked by drones, 😳 there is a hit in a residential building, a house in Mykolaiv is burning...', 'Here’s what we are reading today: Ukraine downed 44 of the 64 Russian missiles and drones that attacked Kyiv, Lviv, Mykolaiv, Dnipropetrovsk and Kharkiv regions on Wednesday morning. The attack took 4 lives and injured over 40 people.', '10:44 am in #Kyiv Morning from Ukraine! The night was calm in Kyiv. But the news from Kharkiv makes me sick and angry. russian drones attacked a gas station in Kharkiv at night, causing an oil spill that, in turn, set 14 private residential buildings on fire. A fire in a…'] |
| Ukraine_War | 9 | 1810 | ['ukraine', 'ukrainians', 'ukrainian', 'russia', 'russians', 'russian', 'country', 'propaganda', 'nazis', 'ethnic'] | ['Ukraine=Russia', 'The is Ukraine', 'Ukraine is Russia...'] |
| Ukraine_War | 11 | 1759 | ['ukraine', 'russia', 'supporting', 'support', 'democracy', 'war', 'country', 'americans', 'america', 'right'] | ['The World can See that the , , and will not support Democracy at home, so we can only Assume this is why they want Ukraine to lose the War!!!!!!!!!!', "And we do and will support Ukraine. Go with Tucker to Russia if you don't want to be an American. But in this country we defend liberal democracy. Isolationism doesn't work. It never has", "If you don't support Ukraine, you support Putin."] |
| Ukraine_War | 17 | 1604 | ['ukraine', 'crimea', 'ukrainian', 'ukrainians', 'putin', 'russia', 'war', 'russian', 'nato', 'nukes'] | ['Then, too, the war in Ukraine would never have started if Russia had not sent weapons of war to Ukraine', "You realize that the only way to prevent a nuclear war is to aid Ukraine, right? It's not a proxy war because we didn't start the war. We can't send our troops there to help because Putin threatened the US if we did. Because it'd be over already if we had put boots on the ground.", 'The war in Ukraine will not end with the defeat of Russia. Putin will not stop. For the war to end, Ukraine will have to give up some land. We can spend 10s of billions of dollars on a war that Ukraine cannot win, or we can push Ukraine to give up some land and end the war.'] |
| Ukraine_War | 20 | 1541 | ['senate', 'bipartisan', 'aid', 'senators', 'senator', 'immigration', 'border', 'israel', 'funding', 'mcconnell'] | ['Senate advances aid bill for Ukraine, Israel and Taiwan without provisions for U.S. border', 'McConnell calls for the Senate to move on from the border security package - and still try to pass aid for Ukraine, Israel, and Taiwan.', 'US Senate blocks Ukraine, Israel aid bill  Senate Republicans voted against the bipartisan border security deal that was part of the $118 billion aid package, which included $60 billion for Ukraine, according to the Hill. Now, the Senate may take up the…'] |
| Ukraine_War | 24 | 1317 | ['aid', 'ukraine', 'funding', 'assistance', 'support', 'fund', 'war', 'give', 'democrats', 'maga'] | ['No. No more aid for Ukraine', "You want us to tell Congress? How about you tell Congress to support the Ukraine aid bill otherwise you'll pull funding. No aid to Ukraine means no aid for you.", 'No more aid to Ukraine'] |
| Ukraine_War | 28 | 1201 | ['ukraine', 'ukrainians', 'ukrainian', 'war', 'kyiv', 'russia', 'russian', 'nato', 'allies', 'troops'] | ['Why the Ukraine War Is Not What You Think', 'Why we should care about it at all ? Main question is how this war is gonna end ? We know what Putin wants, we know what Ukraine wants, but problem is that Ukraine is not gonna win this war, US and EU failed with help . So what is next step ?', 'Things must be going worse than I thought in Ukraine if they are having to make women fight the war. Are all of the men dead? Time to negotiate peace and end this needless war that has killed so many Ukrainians. Americans will no longer be paying for Zelensky’s war.'] |
| Ukraine_War | 40 | 632 | ['putin', 'russian', 'trump', 'vladimir', 'russia', 'biden', 'gop', 'tuberville', 'maga', 'president'] | ['Tommy Tuberville thinks if he stops the US from giving aid to Ukraine, Putin will stop invading Ukraine. I would usually have a difficult time thinking a United States Senator could possibly be this stupid, but it is Tommy Tuberville.', 'Mike Johnson, Who are You working for ? 1. Americans who Pay You ? OR 2. Comrade Putin & Trump ?', "So you're an idiot and a traitor willing to sell the nation out to our enemies to win the next election. Got it, Tuberville. How much did Putin pay you?"] |